\documentclass[aps,nofootinbib,superscriptaddress,preprint]{revtex4-1}

\usepackage{amssymb, amsmath, graphicx}
\usepackage{color}
\usepackage[colorlinks=true, urlcolor=blue, linkcolor=blue,
citecolor=blue, hyperindex=true, linktocpage=true]{hyperref}

\newcommand{\non}{\nonumber \\}

\newcommand{\dd}{\mathrm{d}}
\renewcommand{\d}{\ensuremath{\mathrm{d}}}

\DeclareMathOperator{\sgn}{sgn}

\newcommand{\eps}{\epsilon}

   \newcommand{\ome}{\omega}

    \newcommand{\cD}{{\cal D}}

\newcommand{\CC}{\mathbb{C}}

\newcommand{\RR}{\mathbb{R}}

\newcommand{\re}{\ensuremath{\mathrm{Re}}}
\newcommand{\im}{\ensuremath{\mathrm{Im}}}
\newcommand{\e}{\ensuremath{\mathrm{e}}}

\newcommand{\pa}{\partial}

\newcommand{\rar}{\rightarrow}

\newcommand{\be}{\begin{equation}}
\newcommand{\ee}{\end{equation}}
\newcommand{\bea}{\begin{align}}
\newcommand{\eea}{\end{align}}

\newcommand\lrpar{\raise .8ex\hbox{$^\leftrightarrow$} \hspace{-9pt}
\partial}
\newcommand\lpar{\raise .8ex\hbox{$^\leftarrow$} \hspace{-9pt}
\partial}
\newcommand\rpar{\raise .8ex\hbox{$^\rightarrow$} \hspace{-9pt}
\partial}

\begin{document}

\title{Two-point function of a $d=2$ quantum critical metal in the limit $k_F\rightarrow\infty$, $N_f\rightarrow
  0$ with $N_fk_F$ fixed}

\author{Petter S\"aterskog}
    \email{saterskog@lorentz.leidenuniv.nl}
\affiliation{Institute Lorentz $\Delta$ITP, Leiden University, PO
  Box 9506, Leiden 2300 RA, The Netherlands}
  
\author{Balazs Meszena}
    \email{meszena@lorentz.leidenuniv.nl}
\affiliation{Institute Lorentz $\Delta$ITP, Leiden University, PO
  Box 9506, Leiden 2300 RA, The Netherlands}

\author{Koenraad Schalm}
    \email{kschalm@lorentz.leidenuniv.nl}
\affiliation{Institute Lorentz $\Delta$ITP, Leiden University, PO
  Box 9506, Leiden 2300 RA, The Netherlands}

\begin{abstract}
We show that the fermionic and bosonic spectrum of $d=2$ fermions at finite density coupled to a critical boson can be determined non-perturbatively in the combined
limit $k_F\rar {\infty}$, $N_f \rar 0$ with $N_fk_F$ fixed. In this double
scaling limit, the boson two-point function is corrected, but only at one-loop. This double scaling limit therefore
incorporates the leading effect of Landau damping. The fermion
two-point function is determined analytically in real space and
numerically in (Euclidean) momentum space. The resulting spectrum is
discontinuously connected to the quenched $N_f\rar 0$ result. For
$\omega \rar 0$ with $k$ fixed the spectrum exhibits the distinct non-Fermi-liquid behavior previously surmised from the
RPA approximation. However, the exact answer obtained here shows that
the RPA result does not fully capture the IR of the theory.

\end{abstract}

\maketitle
%{\renewcommand{\baselinestretch}{1}
%\footnotesize
%\tableofcontents
%}
%

\section{Introduction}

The robustness of Landau's Fermi liquid theory relies on the
protected gapless nature of quasiparticle excitations around the Fermi
surface. Wilsonian effective field theory then guarantees that these protected
excitations determine the
macroscopic features of the theory in generic circumstances
\cite{Polchinski, Shankar}. Aside from ordering instabilities, there
is a poignant exception to this general rule. These
are special situations
where the quasiparticle excitations interact with other gapless
states. This is notably so near a symmetry breaking quantum critical
point. The associated massless modes should also contribute to the macroscopic
physics. In $d\geq 3$ dimensions this interaction between Fermi surface excitations
and gapless bosons is marginal/irrelevant
and these so-called quantum critical metals can be addressed in perturbation
theory as first discussed by Hertz and Millis \cite{Hertz,Millis,Loehneysen,Sachdev2}. In 2+1 dimensions, however,
the interaction is relevant and the theory is presumed to flow to a
new interacting fixed point \cite{LeeOrig,OganesyanKivFradOrig,MetznerRoheAndergassenOrig,Sachdev2}. This unknown fixed point has been offered as a
putative explanation of exotic physics in layered electronic materials
near a quantum critical point such as the Ising-nematic transition. 
As a consequence, the deciphering of this fixed point theory  
is one of the major open problems in theoretical condensed matter
physics. There have been numerous earlier studies of Fermi surfaces coupled to gapless bosons but to be able to capture their physics one has almost always been required to study certain simplifying limits \cite{Metlitski1,Metlitski2,Khveshchenko1,Stanford1,Stanford2,Allais:2014fqa,HolderMetzner,Stanford3,Chubukov,Fitzpatrick:2014cfa,Stanford4}.

In this article we show that the fermionic and bosonic spectrum of the most elementary $d=2$
quantum critical metal can be computed
non-perturbatively in the double limit where the Fermi-momentum
  $k_F$ is taken large, $k_F\rightarrow\infty$, while the number of fermion species $N_f$ is
  taken to vanish, $N_f\rightarrow0$, with the combination
  $N_fk_F$ held constant. This is an extension of
  previous work \cite{quenched} where we studied the purely quenched
  limit $N_f\rightarrow0$ {\em followed} by the limit
  $k_F\rightarrow\infty$. In this pure quenched $N_f\rar 0$ limit the boson
  two-point function does not receive any corrections and the fermion
  two point function can be found exactly. However, it is well known
  that for finite $N_f$ and $k_F$ the boson
receives so-called Landau damping contributions that dominate
the IR of the theory. %
These Landau damping corrections are always proportional to $N_f$,
and a subset of these are also proportional to $k_F$. These terms in
particular influence the IR as the large scale, low energy behavior should emerge when $k_F$ is large.
Studying the double scaling limit where the combination
$N_fk_F$ is held fixed gives a more complete understanding of the
small $N_f$ {and/or} large $k_F$ limit {and their interplay}. In particular, 
this new double scaling limit makes precise previous results in the
literature on the RPA approximation together with the {$N_f\rightarrow0$} limit and the strong forward scattering approximation
{\cite{Altshuler1,Altshuler2,Metzner97,ChubukovNew}}.
Importantly, we shall show that the RPA results qualitatively
  capture the low energy at fixed momentum regime, but not the full IR
  of the theory in the double scaling limit.
The idea of this limit is similar to the limit taken in \cite{kf/N} where they study a similar model, but in a matrix large $N$ limit. In this limit they keep the quantity $k_F^{d-1}/N$ fixed while taking both $N$ and $k_F$ large.

All the results here refer to the most elementary quantum critical
metal. This is a set of $N_f$ free spinless fermions at finite density
interacting with a free massless scalar through a simple Yukawa
coupling. Its action reads (in Euclidean time)
\begin{align}
S=\int\!\mathrm{d}x\mathrm{d}y\mathrm{d}\tau\left[\psi_{j}^{\dagger}\left(-\partial_{\tau}+\frac{\nabla^{2}}{2m}+\mu \right)\psi^{j}+\frac{1}{2}\left(\partial_{\tau}\phi\right)^{2}+\frac{1}{2}\left(\nabla\phi\right)^{2}+\lambda\phi\psi_{j}^{\dagger}\psi^{j}\right],\label{action}
\end{align}
where $j=1\ldots N_{f}$ sums over the $N_f$ flavors of fermions and 
$\mu=\frac{k_{F}^2}{2m}$.
We will assume a spherical Fermi surface, meaning $k_F$ both sets the
size of the Fermi surface, $2\pi k_F$, and the Fermi surface
curvature, $1/k_F$. We will study the fermion and boson
  two-point functions of this theory in the double scaling limit $N_f
  \rar 0$, $k_F \rar \infty$. By this we mean that we take
  $k_F\rightarrow\infty$ while keeping the external momenta (measured from Fermi surface), energies,
  the coupling scale $\lambda^2$ and the Fermi velocity $v = k_F/m$
  fixed. We shall not encounter any
  UV-divergences, but to address any ambiguities that may arise the usual
  assumption is made that the above theory is an effective theory
  below an energy and momentum scale $\Lambda_0, \Lambda_k$, each of
  which is already much smaller than $k_F$ ($\Lambda_0,\Lambda_k \ll k_F$). We do not address fermion pairing instabilities in this work. They have been studied and found for similar models in other limits, outside of the particular double scaling limit studied here \cite{kf/N,HeavyFermions,Cooper1,Cooper2,raghuNew}.

\section{Review of the quenched approximation ($N_f\rightarrow0$ first, $k_F\rightarrow\infty$ subsequently)}

Let us briefly review the earlier results of \cite{quenched} as
   they are a direct inspiration for the double scaling
   limit.

Consider the fermion two-point function for the action above,
Eq. \eqref{action}. Coupling the fermions to external sources and
integrating them out, and taking two derivatives w.r.t. the
source, the formal expression for this two point
function is 
\begin{align}
  \label{eq:35}
  G_{\mathrm{full}}(\ome,k) = \langle \psi^{\dagger}(-\ome,-k)\psi(\ome,k) \rangle= \int \cD
  \phi~{\det}^{N_f}(G^{-1}[\phi]) G(\ome,k)[\phi] e^{- \int
    \frac{1}{2}\left(\partial_{\tau}\phi\right)^{2}+\frac{1}{2}\left(\nabla\phi\right)^{2} }
\end{align}
where $G(\ome,k)[\phi]$ is the fermion two-point function in the
presence of a background field $\phi$, defined by
\begin{align}
  \label{eq:36}
  \left(-\partial_{\tau}+\frac{\nabla^{2}}{2m}+\mu
    +\lambda\phi\right)G(t,x)[\phi] = \delta(t-t')\delta^2(x-x')
\end{align}
In the limit $k_F \rar \infty$, {for external momentum $k$ close to the Fermi surface}, we may approximate the derivative part
with {$-\pa_\tau +iv\partial_x$}. The defining equation for the Green's function can
then be solved in terms of a free fermion Green's function dressed
with {the exponential of a linear functional of $\phi$}. In the quenched $N_f \rar 0$
limit this single {exponentially} dressed Green's function can be averaged
over the background scalar with the Gaussian kinetic term.
The result in
real space is again an exponentially dressed free
  Green's function
  \begin{align}
    \label{eq:17}
    G_{R, N_f \rar 0}(r,t) =
    G_{R,\mathrm{free}}(r,t) e^{I(t,r)}
  \end{align}
with the exponent $I(r,t)$ given by
\begin{align}
I(\tau,r)=\lambda^2\int\frac{\d\omega\d k_x\d k_y}{(2\pi)^3}\frac{\cos(\tau\omega-rk_x)-1}{(i\omega-k_x v)^2}G_B(\omega,k)~,\label{expIntDef}
\end{align}
and $r$ conjugate to momentum measured from the Fermi surface ($k_x$), not the
origin. Here $G_B(\omega,k)$ is the free boson Green's function determined by the
explicit form of the boson kinetic term in the action
Eq. \eqref{action}. 
This is of course a known function and due to this simple dressed
expression the retarded Green's function and therefore the fermionic spectrum of this model can be determined exactly in the
limit $N_f \rar 0$. The 
retarded Green's function in momentum space reads \cite{quenched} (here $\omega$ is Lorentzian)
\begin{align}
\label{eq:retarded}
\begin{split}G_{R, N_f \rar 0}(\omega,k_x)=\frac{1}{\omega-k_xv+\frac{\lambda^{2}}{4\pi\sqrt{1-v^{2}}}\sigma(\omega,k_x)}\end{split},
\end{align}
where $\sigma$ is the solution of the equation
\begin{align}
\label{eq:sigmaeq}
\frac{\lambda^{2}}{4\pi\sqrt{1-v^{2}}}(\sinh(\sigma)-\sigma\cosh(\sigma))+v\omega-k_x-\cosh(\sigma)(\omega-k_xv+i\epsilon)=0,
\end{align}
with $k_x$ the distance from the Fermi surface, $v=k_{F}/m$ is
the Fermi velocity, and $\eps \rar 0^+$ is an $i\eps$ prescription
that selects the correct root.

This non-perturbative result already describes interesting singular
fixed point behavior: the spectrum exhibits non-Fermi liquid scaling behavior with multiple
Fermi surfaces \cite{quenched}. Nevertheless, it misses the true IR of
the theory as the quenched limit inherently misses the physics of
Landau damping. This arises from fermion loop corrections to the boson
propagator that are absent for $N_f \rar 0$. 
Below the Landau damping scale $\omega<
\sqrt{\lambda^2 N_f k_F}$ the physics is expected to differ from the
quenched approximation.

\section{Loop-cancellations and boson two-point function}
It is clear from the review of the quenched derivation that
  finite $N_f$, i.e. fermion loop corrections, that only change the boson two-point
  function, can readily be corrected for by replacing the free boson
  two-point function $G_B(\omega, k)$ by the (fermion-loop) corrected boson two-point
  function in Eq. \eqref{expIntDef} (valid at large $k_F$). 
  This is the essence of many RPA-like approximations
  previously studied. A weakness is that finite $N_f$ corrections will 
  also generate higher-order boson interactions and these can invalidate
  the simple dressed expression obtained here.     

At the same time, it has been known for some time that finite density fermion-boson models with simple Yukawa
scalar-fermion-density interactions as in Eq.~\eqref{action} have
considerable cancellations in fermion loop diagrams for low energies
and momenta after symmetrization
\cite{Metzner97,Neumayr:1998hj,Kopietz}. These cancellations make loops with
more than three interaction vertices
$V\geq 3$ finite as the external momenta and energies are scaled uniformly to
zero. We will now argue that this result also means that in the $N_f \rightarrow 0$, $k_F \rightarrow \infty$
  limit with $N_fk_F$ fixed, these $V\ge3$ loops vanish. In this limit {\em
  only} the boson two-point function is therefore corrected and only
at one loop and we can directly deduce that in this double limit the
exact fermion correlation function is given by the analogue of the
dressed Green's function in Eq. \eqref{eq:17}. {We comment on the limitations of considering this limit for subdiagrams in perturbation theory later in this section.}

\begin{figure}
\includegraphics[width=\textwidth]{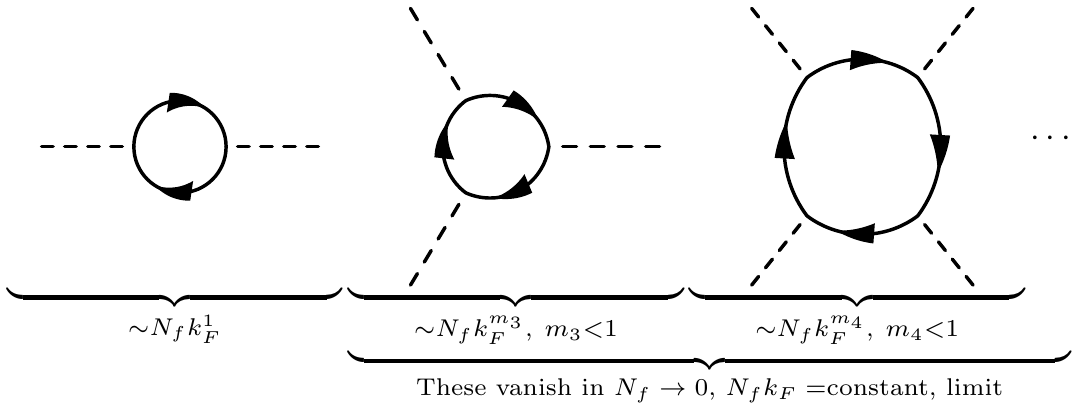}
\caption{\label{fig:fermionLoops}Here we show the dominant scaling of fermion loops with different numbers of vertices in the limit of $N_f\rightarrow0$ with $N_fk_F$ constant. This is the scaling after symmetrizing the external momenta. The two-vertex loop on the left does not get symmetrized and is the only loop that does not vanish in this limit.}
\end{figure}

Consider the quantum critical metal before any approximations; i.e. we have a fully rotationally invariant Fermi surface with a finite $k_F$. The Yukawa coupling shows that the boson couples to the density operator $\psi^\dagger(x)\psi(x)$. All corrections to the boson therefore come from fermionic loops with fermion density vertices. These loops always show up symmetrized in the density vertices. Consider a fermion loop with a fixed number V of such density
vertices, dropping the overall coupling constant dependence, and arbitrary incoming energies and momenta. Such a loop (ignoring the overall momentum conserving $\delta$-function) has energy dimension $3-V$. These fermionic loops are all UV finite so they are independent of the scale of the UV cut-offs. There are only two important scales, the external bosonic energies and momenta $\omega_i, k_i$, the fermi momentum $k_F$.
A symmetrized $V$-point loop can by dimensional analysis be written as
\begin{align}
I(\{\omega_i\},\{k_i\})=k_F^{3-V}f(\{\omega_i/k_F\},\{k_i/k_F\})%\sum_{j}k_F^{n_j} \omega_1^{3-V-n_j} f_j(\{\omega_i/\omega_1\},\{k_i/\omega_i\})
\end{align}
Since fermion loops of our theory with $V\ge3$ vertices have been shown to be finite as external energies and momenta are uniformly scaled to 0 \cite{Neumayr:1998hj}, we thus have that $f$ is finite as $k_F$ is taken to infinity. This in turn means that $I(\{\omega_i\},\{k_i\})$ scales as $k_F^n$ with $n\leq0$ for large $k_F$ when $V\ge3$.  \footnote{Naively corrections to the boson would be expected to scale as $k_F$ since it receives corrections from a Fermi surface of size $2\pi k_F$. However $k_F$ also sets the curvature of the Fermi surface and for a large $k_F$ we approach a flat Fermi surface for which $V\ge3$-loops completely cancel. This is shown in more detail in Appendix \ref{sec:mult-canc}.} Note that the use of the small external energies and momenta limit from \cite{Neumayr:1998hj} was merely a way of deducing the large $k_F$ limit. We do not rely on the physical IR scaling to be the same as in \cite{Neumayr:1998hj}, indeed we will find it not to be the same. All single fermionic loops additionally contain a sum over fermionic flavors so are therefore proportional to $N_f$. Combining this we see that a fermionic loop with $V\geq3$ density vertices comes with a factor of $N_fk_F^{m_V}$ where $m_V\leq0$ after symmetrizing the vertices. By now considering the combined limit of $N_f\rightarrow0$ and $k_F\rightarrow\infty$ with $N_fk_F$ constant we see that these $V\geq3$ loops all vanish. See Figure \ref{fig:fermionLoops}.\\
We have now concluded that for a fixed set of external momenta, all symmetrized fermion loops vanish in our combined limit, except the $V=2$ loop. %One may still be worried that 
There is still a possibility that diagrams containing $V>2$ loops are important when taking the combined limit \emph{after} performing all bosonic momentum integrals and summing up the infinite series of diagrams.
%the have a different scaling in $k_F$ after the bosonic momentum integrals have been performed.
In essence, the bosonic integrals and the infinite sum of perturbation theory need not commute with the combined $N_f\rightarrow0$, $k_F\rightarrow\infty$ limit. What the IR of the full theory (finite $N_f$) looks like is not known so taking the $N_f\rightarrow0$ limit last is currently out of reach. In \cite{Thier} the authors show that divergence of fermionic loops does not cancel under a non-uniform low-energy scaling of energies and momenta where the momenta are additionally taken to be increasingly collinear. The scaling they use is motivated by the perturbative treatment in \cite{Metlitski1} and if this is the true IR scaling and it persists at small $N_f$, then there will be effects unaccounted for in the above.

Regardless of the above mentioned caveat, in keeping the $V=2$ fermion loops we take the combined limit \emph{after} performing the fermionic loop integrals and thus move closer than in our previous work \cite{quenched} to the goal of understanding the IR of quantum critical metals.
% In essence the UV-limit of the bosonic integrals and the $k_F\rightarrow\infty$ limits need not commute. It is known that these integrals are divergent in the UV for higher order diagrams \cite{Metlitski1,Metlitski2,HolderMetzner}. However, we have assumed cut-offs $\Lambda_0, \Lambda_k$ for the boson and these are kept fixed in the large $k_F$ limit. No UV-limit is therefore taken for the boson momentum integrals and they necessarily commute with our combined limit.
To summarize: the ordered limit we consider is
\begin{enumerate}
\item{We first perform rotationally invariant finite $k_F$ fermionic loop integrals.}
\item{Then we take the limit $N_f\rightarrow0$, $k_F\rightarrow\infty$ with $N_fk_F$ fixed; this only keeps $V=2$ loops.}
\item{Next we perform bosonic loop integrals.}
\item{Finally we sum all contributions at all orders of the coupling constant.}
\end{enumerate}
The result above means the fully quantum corrected boson remains Gaussian in this ordered limit and only receives corrections from the $V=2$ loops.

\subsection{Boson two-point function}

We now compute the one-loop correction to the boson two-point function; in {our ordered} double scaling limit this is all we need. We then
substitute the Dyson summed one-loop corrected boson two-point
function into the dressed fermion Green's function to obtain the exact fermionic 
spectrum.

The one-loop correction---the boson polarization---in the double
scaling limit is given by the large $k_F$ limit of the two-vertex fermion loop. This can be calculated using a linearized fermion dispersion:
\begin{align}
\Pi_1\left(Q\right)=\lambda^{2}N_{f}k_{F}\int\! \frac{\dd k_{0}\dd k d\theta}{(2\pi)^{3}}\frac{1}{\left(ik_{0}-vk\right)\left(i\left(k_{0}+q_{0}\right)-v\left(k+|\vec{q}|\cos\theta\right)\right)}.
\end{align}
Note from the $\cos\theta$ dependence in the numerator that we are not making a ``patch'' approximation. In the low energy limit this angular dependence is the important contribution of the rotationally invariant fermi-surface, whereas the subleading terms of the dispersion can be safely ignored. As stated earlier, the result of these integrals is finite. However,
it does depend on the order of
integration. The difference is a constant $C$
\begin{align}
\Pi_{1}\left(Q\right)&=\frac{\lambda^{2}N_fk_F}{{2\pi}v}\left(\frac{|q_{0}|}{\sqrt{q_{0}^{2}+v^{2}\vec{q}^{2}}}+C\right)\nonumber\\
&\equiv M_D^2\left(\frac{|q_{0}|}{\sqrt{q_{0}^{2}+v^{2}\vec{q}^{2}}}+C\right).\label{eq:1-loop}
\end{align}
As pointed out in for instance \cite{Stanford4,Fitzpatrick:2014cfa},
the way to think about this ordering ambiguity is that 
one should strictly speaking first regularize the theory and introduce a one-loop
counterterm. This counterterm has a finite ambiguity that needs to be
fixed by a renormalization condition. Even though the loop momentum
integral happens to be finite in this case, the finite counterterm
ambiguity remains. The correct renormalization condition 
is the choice $C=0$. This choice corresponds to
the case when the boson is tuned to criticality since a non-zero $C$
would mean the presence of an effective mass generated by quantum
effects. 

A more physical way to think of the ordering ambiguity is as 
the relation between the frequency ($\Lambda_{0}$)
and momentum ($\Lambda_{k}$) cutoff. 
We will assume that $\Lambda_{k}\gg\Lambda_{0}$ --- which
means that we evaluate the $k$ integral first and then the frequency
$k_{0}$ integral. In this case $C=0$ directly follows.

\section{Fermion two-point function}
{%
With the single surviving one-loop correction to the boson two-point function in hand, we can immediately write down the expression for the full
fermion two-point function. This is the same dressed expression Eq. \eqref{expIntDef} as in \cite{quenched} but with a modified boson propagator $G_B=1/(G_{B,0}^{-1}+\Pi)$, with $\Pi$ the one-loop polarization of
Eq. \eqref{eq:1-loop}}. Substituting this in we thus need to calculate the integral
\begin{align}
I(\tau,r)=\lambda^2\int\frac{\d\omega\d k_x\d k_y}{(2\pi)^3}\frac{\cos(\tau\omega-rk_x)-1}{(i\omega-k_x v)^2\Big(\omega^2+k_x^2+k_y^2+M_D^2\frac{|\omega|}{\sqrt{v^2(k_x^2+k_y^2)+\omega^2}}\Big)}\label{messy}.
\end{align}

At this moment, we can explain clearly how our result connects to previous approaches. A similarly dressed propagator can be proposed based on extrapolation from 1d results \cite{Altshuler1,Metzner97}.
An often used approximation in the literature is to now study this below the scale $M_D$, see e.g \cite{Altshuler2,Sachdev2}. This is the physically most interesting limit since in the systems of interest $N_f$ is order one and we are considering large $k_F$. In this limit the
polarisation term will dominate over the kinetic terms, but since the rest of the integrand in (\ref{messy})
has no $k_y$ dependence, it is necessary to keep the $k_y$
term in the boson propagator. The $\omega$ and $k_x$ momenta will
suppress the integrand when they are of order $\lambda^2$ whereas the
$k_y$ term will do so once it is of order
$\lambda^{2/3}M_D^{2/3}$. This means that for $M_D\gg\lambda^2$, the relevant $k_y$ will be much larger than the relevant $\omega$ and $k_x$. This argues that we can truncate to the large $M_D$ propagator
\begin{align}
G_{B,M_D\rightarrow\infty}(\omega,k_x,k_y)=\frac{1}{k_y^2+M_D^2\frac{|\omega|}{v|k_y|}}.
\end{align}
This Landau-damped propagator has been used extensively, for instance 
\cite{Sachdev2,Altshuler2}. In \cite{Altshuler2} this propagator was %
used for the type of non-perturbative calculation we are proposing
here. We discuss this here, as we will now show that using this simplified propagator has a problematic feature. This propagator
only captures the leading large $M_D$ contribution but the
non-perturbative exponential form of the exact Green's function sums up powers of the propagator which then
are subleading in $M_D$.

Using the large $M_D$ truncated boson Green's function the integral $I(\tau, r)$ to be evaluated simplifies to
\begin{align}
I_{M_D\rightarrow\infty}=\lambda^2\int\frac{\d\omega\d k_x\d k_y}{(2\pi)^3}\frac{\cos(\tau\omega-xk_x)-1}{(i\omega-k_x v)^2\Big(k_y^2+M_D^2\frac{|\omega|}{v|k_y|}\Big)}.
\end{align}
Writing the cosine in terms of exponentials we can perform the $k_x$
integrals using residues by closing the contours in opposite half
planes. Summing up the residues gives
\begin{align}
I_{M_D\rightarrow\infty}=-\lambda^2\int\d\omega\d k_y\frac{k_y^2 |r| e^{-\left| \omega \right|   \left(\frac{|r|}{v}+i\sgn(r) \tau
   \right)}}{8 \pi ^2 M_D^2 \left|  \omega k_y \right| +k_y^4 v}.
\end{align}
The $k_y$ integral can be performed next to yield
\begin{align}
I_{M_D\rightarrow\infty}=-\lambda^2\int\d\omega\frac{  \left| r\right|  e^{-\left| \omega \right|   \left(\frac{|r|}{v}+i\sgn(r) \tau
   \right)}}{12 \sqrt{3} \pi (M_D^2v^5\left| \omega \right|)^{1/3} }.
\end{align}
The primitive function to this $\omega$ integral is the upper
incomplete gamma function, with argument $2/3$. Evaluating this
incomplete gamma function in the appropriate limits and substituting
the final expression for $\underset{M_D\rar \infty}{I}(\tau,r)$
into the expression for the fermion two-point function gives us:
\begin{align}
\label{eq:31}
\underset{M_D\rightarrow\infty}{G_f}(\tau,r)=\frac{1}{2\pi(ir-v\tau)}\exp\left(-\frac{\left| r\right| }{l_0^{1/3} \left(\left| r\right|+iv\sgn(r) \tau\right)^{2/3}}\right)
\end{align}
where the length scale $l_0$ is given by
\begin{align}
l_0^{1/3}=\frac{6\sqrt{3} \pi v M_D^{2/3} }{\Gamma \left(\frac{2}{3}\right)\lambda^{2}}.
\end{align}

\begin{figure}
\centering{}
\includegraphics{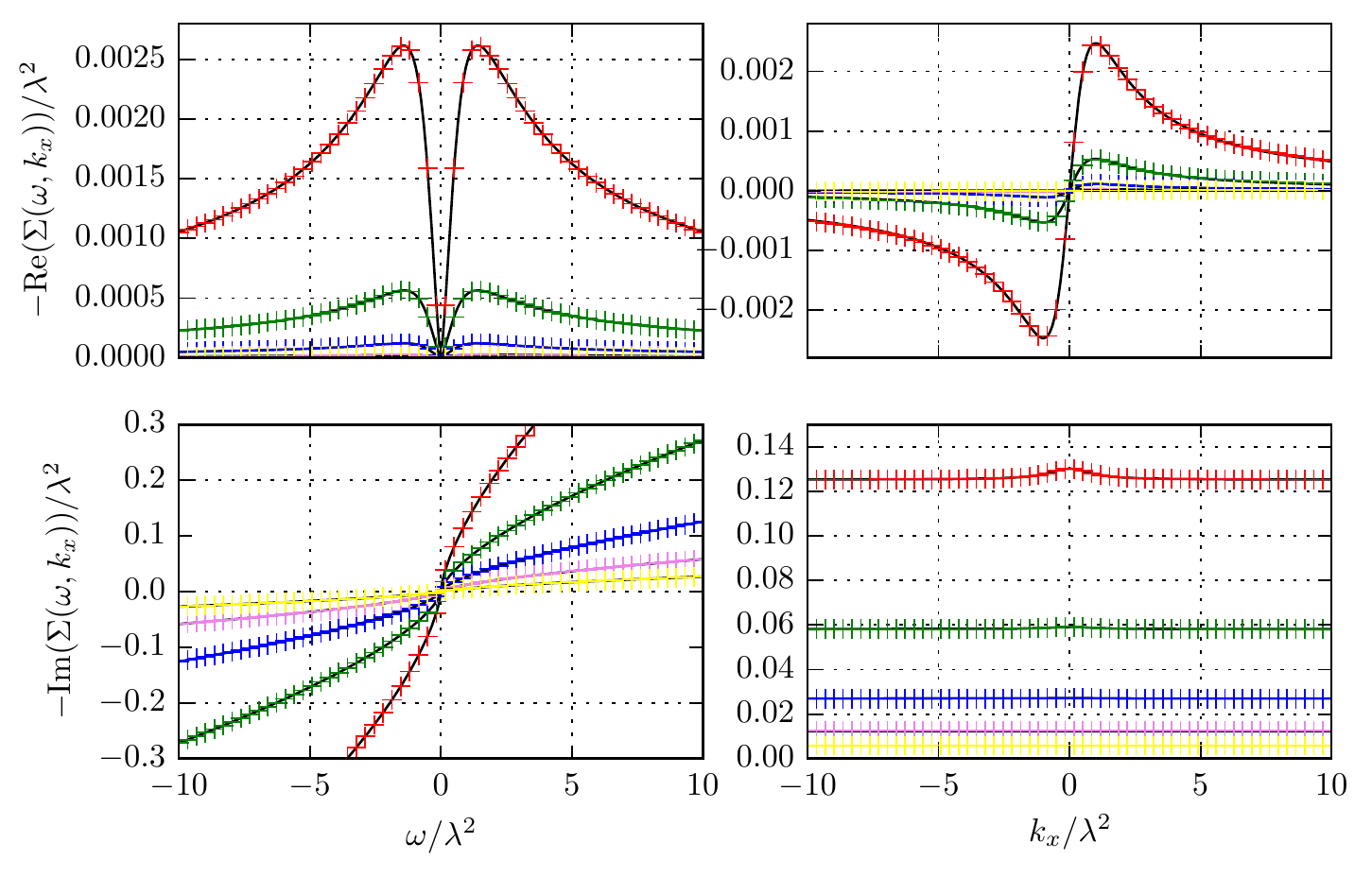}
\caption{Real and imaginary parts of the self energy obtained using the large-$k_F$ Landau-damped propagator. This plot shows the agreement between the numerics and the analytical solution, verifying that both solutions are correct. Notice the difference in magnitude between the real and imaginary part. The agreement of the real parts shows that the numerical procedure has a very small relative error. All plots are for the $k_x,\omega=\lambda^2$ slice with $v=1$.\label{fig:benchmark}}
\end{figure}

This result has been found earlier in \cite{Altshuler1} (see also
\cite{Metzner97}). However, this real space expression hides the
inconsistency of the approach. This becomes apparent in its momentum space representation.
The Fourier transform of the real space Green's function
\begin{align}
\underset{M_D\rightarrow\infty}{G_f}(\omega,k)=\int\d\tau\d r\frac{\e^{i(\omega\tau-kr)}}{2\pi(ir-v\tau)}\exp\left(-\frac{\left| r\right| }{l_0^{1/3} \left(\left| r\right|+iv\sgn(r) \tau\right)^{2/3}}\right)
\label{cheatIntDef}
\end{align}
is tricky, but remarkably can be done exactly. We do so in appendix
\ref{sec:ferm-greens-funct}. The result is
\begin{align}
\begin{split}
\underset{M_D\rightarrow\infty}{G_f}(\omega,k_x)=
&\frac{1}{i\omega-k_x v}  \cos \left(   \frac{\omega}{v l_0^{1/2}(\omega/v+ik_x)^{3/2}}      \right)\\
&+\frac{   6\sqrt{3}  i  \Gamma \left(\frac{1}{3}\right) \omega^{2/3} }{8 \pi l_0^{1/3}v^{5/3}  (\omega/v+ik_x)^2} \,
   _1F_2\left(1;\frac{5}{6},\frac{4}{3};  - \frac{\omega^{2}}{4l_0v^2(\omega/v+ik_x)^3}   \right) +\\
&+
       \frac{3  \sqrt{3}  i   \Gamma \left(-\frac{1}{3}\right)  \omega^{4/3}  }{8 \pi l_0^{2/3}v^{7/3} (\omega/v+ik_x)^3}
 \,
   _1F_2\left(1;\frac{7}{6},\frac{5}{3}; - \frac{\omega^{2}}{4l_0v^2(\omega/v+ik_x)^3}  \right) .
   \end{split}
   \label{eq:GlargeNfAppEucResult1}
\end{align}
This expression has been compared with numerics to verify its correctness; see Fig. \ref{fig:benchmark}. 

\begin{figure}
\centering{}
\includegraphics{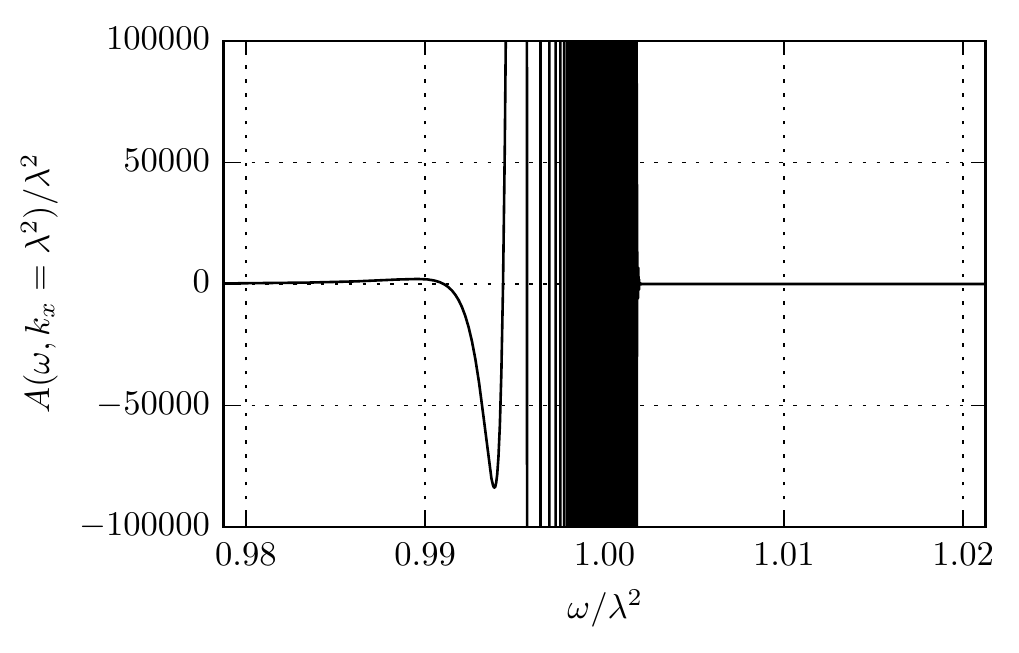}
\caption{Exact fermion spectral function based on the large-$M_D$
  approximation for the exact boson propagator. Notice that the function is not positive everywhere. Here $k=\lambda^2$, $v=1$ and $M_D=2\pi\lambda^2$.\label{fig:inconsistency}}
\end{figure}

We can now show the problematic feature. Recall that Eq. \eqref{eq:GlargeNfAppEucResult1} is the Green's
function in Euclidean signature. Continuing to the imaginary line,
$\omega=-i\omega_R$, this becomes the proper retarded Greens function,
$G_R(\omega_R,k_x)$, and from this we can obtain the spectral function
$A(\omega_R, k_x)=-2\im\ G_R(\omega_R, k_x)$. As it encodes  the
excitation spectrum, the spectral function ought to be a
\emph{positive} function that moreover equals $2\pi$ when integrated over all energies $\omega_R$, for any momentum $k$.
This large $M_D$ spectral function
contains an oscillating singularity at $\omega_R=vk_x$. %
We are free to move the contour into complex
$\omega_R$-plane by deforming $\omega_R\rightarrow\omega_R+i\Omega$
where $\Omega$ is positive but otherwise arbitrary. Upon doing this it
is easy to numerically verify that indeed the integral over
$\omega_R$ gives $2\pi$. However, if we look at the behaviour close
to the essential singularity, the function oscillates rapidly and does
not stay positive as one approaches the singularity; see
Fig. \ref{fig:inconsistency}. This reflects that the large $M_D$ approximation done in this way is not consistent.  %
Even though the approximation for the
exponent $I(\tau,r)\equiv\frac{\tilde{I}(\tau,r)}{(M_D)^{2/3}}$ is valid to leading order in $1/M_D$, this
is not systematic after exponentiation to obtain the fermion two-point function
\begin{align}
\underset{M_D\rightarrow\infty}{G_f}(\tau,r)=\frac{1}{2\pi(ir-v\tau)}\exp\left(
\frac{\tilde{I}(\tau,r)}{
M_D^{2/3}}
+\mathcal{O}\Big(\frac{1}{ (M_D)^{4/3}}\Big)\right).
\end{align}
Reexpanding the exponent one immediately sees that keeping only the
leading term in $I(\tau,r)$ mixes at higher order with the subleading
terms at lower order in $1/M_D$
\begin{align}
\underset{M_D\rightarrow\infty}{G_f}(\tau,x)=\frac{1}{2\pi(ir-v\tau)}\left(1+
\frac{\tilde{I}(\tau,r)}{
M_D^{2/3}}
+\mathcal{O}(M_D^{-4/3})
+\frac{1}{2}
\left(
\frac{\tilde{I}(\tau,r)}{
M_D^{2/3}}
+\ldots\right)^2
\right).
\end{align}
Despite this problematic feature, we will show from the exact result that in the IR $\underset{M_D\rightarrow\infty}{G_f}$ (with a small modification) does happen to capture the correct physics.

\subsubsection{{Exact fermion two-point function for large $k_F$ with
  $M_D$ fixed; $v=1$}}

We therefore make no further assumption regarding the value of $M_D$ and we return to the full integral Eq. \eqref{messy} to
determine the real space fermion two-point function. Solving this in
general is difficult, and to simplify mildly we consider the special
case $v=1$. In our previous studies of the quenched {$M_D=0$} limit
we saw that this choice for value of $v$ is actually not very special, even though it
appears that there is an enhanced symmetry. In fact, nothing abruptly
happens as $v\rightarrow1$, except that the quenched $M_D=0$ solution can be
written in closed form for this value of $v=1$. Nor for the case of
large $M_D$ is the
choice $v=1$ in any way special. As can be seen above in Eq.~\eqref{eq:31}
  for large $M_D$ all $v$ are equivalent up to a rescaling of $\tau$
  versus $r$ and a rescaling of the single length scale $l_0$. We may
  therefore expect that for a finite $M_D$, the physics of
  $0<v<1$ is qualitatively the same as the (not-so-) special case
  $v=1$. 

After setting $v=1$ and changing to spherical coordinates we have
\begin{align}
\label{eq:32}
I=\lambda^{2}\int\d \tilde{r}\d\phi\d\theta\e^{2i\phi}\frac{\cos\big(\tilde{r}\sin(\theta)(\tau\sin(\phi)-r\cos(\phi))\big)-1}{8\pi^3\sin(\theta)^2\big(M_D^2|\sin(\phi)|+\tilde{r}^2/\sin(\theta)\big)}~.
\end{align}
Performing the $\tilde{r}$ integral gives us
\begin{align}
I=\pi^4\lambda^2\int\d\phi\d\theta\e^{2i\phi}\frac{\e^{-M_D|\tau\sin(\phi)-r\cos(\phi)|\sqrt{|\sin(\phi)|\sin(\theta)^3}}-1}{16M_D\sqrt{|\sin(\phi)|\sin(\theta)^3}}~.
\end{align}
Note that if the signs of both $\tau$ and $r$ are flipped, then this
is invariant. Changing the sign of only $\tau$, and simultaneously
making the change of variable $\phi\rightarrow-\phi$, then the (real)
fraction is invariant but the exponent in the prefactor changes
sign. Thus, $I$ goes to $I^*$ as the sign of either $\tau$ or $r$ is
changed. Without loss of generality, we can assume that both of them
are positive from now on. We further see that the integrand is
invariant under $\phi\rightarrow\phi+\pi$, so we may limit the range
of $\phi$ to $(0,\pi)$ by doubling the value of integrand. Similarly
we limit $\theta$ to $(0,\pi/2)$ and multiply by another factor of 2. We then make the changes of variables:
\begin{align}
\begin{split}
\phi&=\tan^{-1}(s)+\pi/2\\
\theta&=\sin^{-1}(u^{2/3})
\end{split}
\end{align}
with $s \in \RR$ and $u$ is integrated over the range $(0,1)$. 
For convenience we introduce the function
\begin{align}
z(s)=M_D|sr+\tau|(1+s^2)^{-3/4}.
\end{align}
Now the two remaining integrals can be written as
\begin{align}
I=\frac{\lambda^2}{M_D}\int\d s\d u\frac{(\e^{-uz(s)}-1)(s-i)}{6\pi^2(s+i)(1+s^2)^{3/4}u^{4/3}\sqrt{(1-u^{4/3})}}.
\label{eqn:usIntegralWithz}
\end{align}
After expanding the exponential we can perform the $u$ integral term by term. We are left with
\begin{align}
I=\frac{\lambda^2}{M_D}\int\d s\sum_{n=1}^\infty\frac{(1+s^2)^{1/4}(-z(s))^n}{8\pi^{3/2}n!(i+s)^2}\frac{\Gamma\Big(\frac{3n-1}{4}\Big)}{\Gamma\Big(\frac{3n+1}{4}\Big)}.
\end{align}
This can be resummed into a sum of generalized hypergeometric
functions, but this is not useful at this stage. Instead we once again integrate term by term. %
Collecting the prefactors and introducing the constant $a=\tau/r$, the $n$-th term can be written as
\begin{align}
I=\sum_{n=1}^\infty c_n \int\d s\,
(s-i)^2 |s+a|^n (1 + s^2)^{- (7 + 3 n)/4}.
\end{align}
This can be written as
\begin{align}
I=\sum_{n=1}^\infty c_n \int\d s\d w\,
(s-i)^2 |s+a|^n \frac{\e^{-w (s^2 + 1)} w^{3 (1+n)/4}}{(3(1+n)/4)!},
\end{align}
where $w$ is integrated on $(0,\infty)$. After splitting the integral
at $s=-a$ to get rid of the absolute value we can calculate the $s$
integrals in terms of confluent hypergeometric functions
$_1F_1(a,b;z)$. Adding the two halves $s<-a$ and $s>-a$ of the integral we have
\begin{align}
\begin{split}
I=&\sum_{n=1}^\infty c_n \int\d w \frac{\Gamma\Big(\frac{1+n}{2}\Big)\e^{-(1+a^2)w}}{2(3(1+n)/4)!}\Bigg(
2w(i+a)^2\ _1F_1\Big(\frac{2+n}{2},\frac{1}{2};a^2w\Big)\\
&+(2+n)\ _1F_1\Big(\frac{4+n}{2},\frac{1}{2};a^2w\Big)
-4aw(2+n)(i + a)\ _1F_1\Big(\frac{4+n}{2},\frac{3}{2};a^2w\Big).
\Bigg)
\end{split}
\end{align}
It may look like we have just exchanged the $s$-integral for the
$w$-integral, but by writing the hypergeometric functions in series form,
\begin{align}
\begin{split}
I=&\sum_{n=1}^\infty c_n \int\d w\sum_{m=0}^\infty
\frac{2^{2 m-1} a^{2 m} \e^{-(1 + a^2) w} w^{
 \frac{n-3}{4} + 
  m}  \Gamma\Big(\frac{1 + n}{2} + m\Big)}{
\Gamma\Big(\frac{7+3n}{4}\Big) \Gamma\Big(2 + 2 m\Big)}\times\\
&\times(n (1 + 2 m - 4 a (i + a) w) + (1 + 2 m) (1 + 2 m - 
      2 (1 + a^2) w))~,
\end{split}
\end{align}
the $w$ integral can now be performed. The result is
\begin{align}
\begin{split}
I=\sum_{n=1,m=0}^\infty
&c_n\frac{(a+i) 4^{m-1} a^{2 m} \left(a^2+1\right)^{-\frac{1}{4} (4 m+n+5)}  \Gamma \left(m+\frac{n+1}{4}\right) \Gamma
   \left(m+\frac{n+1}{2}\right)}{\Gamma (2 m+2) \Gamma \left(\frac{7+3 n}{4}\right)}\times\\
   &\times\left(a \left(2m-6 m n-2 n^2-n+1\right)-i (2 m+1) (n+1)\right).
      \end{split}
\end{align}
The sum over $m$ can be expressed in terms of the ordinary hypergeometric function, $_2F_1(a_1,a_2;b;z)$:
\begin{align}
I=\sum_{n=1}^\infty
&c_n
\frac{(n+1) \left(a^2+1\right)^{-\frac{n}{4}-\frac{1}{4}} \Gamma \left(\frac{n+1}{4}\right)
   \Gamma \left(\frac{n+1}{2}\right) }{24
   (a-i)^2 (a+i) \Gamma \left(\frac{3 n}{4}+\frac{7}{4}\right)}\times\\
   &\times
   \left(a^2 (n+1) (-3 a n+a-i (n+1)) \,
   _2F_1\Bigg(\frac{n+3}{2},\frac{n+5}{4};\frac{5}{2};\frac{a^2}{a^2+1}\right)+\\
   &-6
   \left(a^2+1\right) (a (2 n-1)+i) \,
   _2F_1\left(\frac{n+1}{4},\frac{n+1}{2};\frac{3}{2};\frac{a^2}{a^2+1}\right)\Bigg).
\end{align}
The space-time dependence in this expression is in $a=\tau/r$
and with additional $r$-dependence in the coefficients $c_n$ 
The result above is the value for both $\tau$ and $r$ positive. Using
the known symmetries presented above, the solution can be extended to
all values of $\tau$ and $r$ by appropriate absolute value signs. 
Then changing variables to
\begin{align}
\begin{split}
\tau&=R\cos(\Phi) \\
r&=R\sin(\Phi) 
\end{split}
\end{align}
we have
\begin{align}
\begin{split}
I=\frac{\lambda^2 f\left(R M_D,\Phi\right)}{M_D}
\end{split}
\end{align}
with the function $f(\tilde{R},\Phi)$ given by
\begin{align}
\begin{split}
f(\tilde{R},\Phi)=&\sum_{n=1}^\infty f_n \tilde{R}^n\\
f_n=&
\frac{e^{i \Phi }2^{-1-n}  (-1)^n \Gamma
   \left(\frac{n+1}{4}\right) \left| \sin (\Phi )\right| ^{\frac{1+3 n}{2}}
  }{9 \pi (3 n-1)\Gamma \left(\frac{n}{2}+1\right) \Gamma
   \left(\frac{1+3 n}{4}\right)}\cdot\\
   &\cdot \Bigg(\, _2F_1\left(\frac{n+3}{2},\frac{n+5}{4};\frac{5}{2};\cos
   ^2(\Phi )\right)(n+1)\cdot\\&\ \ \ \ \ \ \cdot \cos ^2(\Phi )  ((1-3n)\cos (\Phi )-i (n+1) \sin (\Phi )) \\
   &\ \ + \,_2F_1\left(\frac{n+1}{4},\frac{n+1}{2};\frac{3}{2};\cos ^2(\Phi )\right) 6((1-2 n) \cos (\Phi
   )-i \sin (\Phi ))\Bigg).
\end{split}
\label{eqn:sum}
\end{align}
This exact infinite series expression for the exponent $I(\tilde{R},\Phi)$ gives us the exact
fermion two-point function in real (Euclidean) space (time).
We have not been able to find a closed form expression for this final
series. Note that $f_n\sim1/n!$ for large $n$, and the series
therefore converges rapidly. Moreover, numerically the hypergeometric
functions are readily evaluated to arbitrary precision (e.g. with
Mathematica), and therefore the value of $f(\tilde{R},\Phi)$ can be robustly evaluated
to any required precision. 

As a check on this result, we can compare it to the exact result in
the quenched $M_D=0$ limit in \cite{quenched}, where the exact answer
was found in a different way. In the limit where $M_D\rightarrow0$
we see that only the first term of this series gives a contribution
and the expression for the exponent collapses to 
\begin{align}
\lim_{M_D\rar 0} I(R,\Phi) = \lambda^2f_1R=\lambda^2\frac{e^{2 i \Phi }}{12 \pi }R.
\end{align} 
In Cartesian coordinates this equals 
\begin{align}
\lim_{M_D\rar 0} I(\tau,r) =\lambda^2\frac{(\tau +i r)^2}{12 \pi  \sqrt{\tau ^2+r^2}}.
\end{align}
This is the exact same expression as found in \cite{quenched} for $v=1$.

There is one value of the argument for which $f(\tilde{R},\Phi)$ drastically simplifies.
For $r=0$ ($\Phi=0,\pi$) we have
\begin{align}
\begin{split}
f_n(\Phi=0)=&-\frac{ (-1)^{n}}{6\pi \Gamma (n+2)}
\end{split}
\label{eqn:sum1}
\end{align}
and thus
\begin{align}
\begin{split}
f(\tilde{R},\Phi=0)=&\frac{1 }{6 \pi }+\frac{e^{- \tilde{R}}-1}{6 \pi  \tilde{R}}.
\end{split}
\label{eqn:sum2}
\end{align}
Further numerical analysis shows that the real part of $f(\tau,r)$
is maximal for $r=0$.%

\subsubsection{The IR limit of the exact fermion two-point function
 compared to the large-$M_D$ expansion}

With this exact real space answer, we can now reconsider why the large
$M_D$ (large $N_fk_F$) limit fails and which expression does reliably capture the
strongly coupled IR physics of interest. 
The expression obtained above, Eq. \eqref{eqn:sum}, is not very useful
for extracting the IR Green's function or the Green's function at a large $M_D$ as the
expression is organized in an expansion around $RM_D=0$.
To study the limit where $RM_D \gg 1$ we can go back to
Eq. \eqref{eqn:usIntegralWithz}. With this expression we see that the exponential in the
integrand, $e^{-uz(s)}$ with $z \sim
M_D|sr+\tau| \sim \tilde{r}$, is generically suppressed for large
$\tilde{R}=RM_D$.  The exceptions are when either $sr+\tau$
is small, $s$ is large, or $u$ is small. The first two cases are also
unimportant in the $\tilde{R} \gg 1$ limit. In the first case we restrict
the $s$ integral to a small range of order $1/\tilde{R}$ around
$-\tau/r$; this contribution therefore becomes more and more negligible
in the limit $\tilde{R} \gg 1$. In the second case we will have a
remaining large denominator in $s$ outside the exponent that also
suppresses the overall integral. Thus for large $\tilde{R}$, the only
appreciable contribution of the exponential term to the integral
in $I(\tau,r)$ arises when $u$ is small. To use this, we first write the integral as
\begin{align}
I_{\mathrm{IR}} &= I_{\mathrm{IR,exp}}+I_{\mathrm{IR},-1}, \\
I_{\mathrm{IR,exp}}(\tau,r)&=\lambda^2\int_{-\infty}^{\infty}\d
 s\!\frac{s-i}{6\pi^2(s+i)(1+s^2)^{3/4}M_D}\left(\int_0^1\d
 u\frac{\e^{-4uz(s)}-1}{u^{4/3}\sqrt{1-u^{4/3}}} -\int_1^\infty\d u\frac{1}{u^{4/3}}\right)\non
 &\simeq \lambda^2\int_{-\infty}^{\infty}\d
  s\!\frac{s-i}{6\pi^2(s+i)(1+s^2)^{3/4}M_D}
\left(\int_0^1\d
  u\frac{\e^{-4uz(s)}-1}{u^{4/3}} -\int_1^\infty \d u \frac{1}{u^{4/3}}\right),\non
 I_{\mathrm{IR,-1}}(\tau,r)&=\lambda^2\int_{-\infty}^{\infty}\d
  s\!\frac{s-i}{6\pi^2(s+i)(1+s^2)^{3/4}M_D}\left(\int_1^{\infty}\d
  u\frac{1}{u^{4/3}}+\int_0^1\d u\left(\frac{-u^{-4/3}}{\sqrt{1-u^{4/3}}}+u^{-4/3}\right)\right)\nonumber.
\end{align}
We have added and subtracted an extra term to each to ensure convergence of
each of the separate terms.
Since the important contribution to $I_{\mathrm{IR,exp}}$
is from the small $u$ region we can extend its range from
(0,1) to $(0,\infty)$. This way, the integrals can then be done 
\begin{align}
I_{\mathrm{IR},\mathrm{exp}}&=\int_{-\infty}^{\infty}\d s\frac{-\lambda^{2}\left| s r+\tau \right|^{1/3}}{
   3^{3/2} \pi  M_D^{2/3} (s+i)^2
   \Gamma \left(\frac{4}{3}\right)}\non
   &=-\frac{\Gamma \left(\frac{2}{3}\right)\lambda^{2}\left|
   r\right| ^{1/3}}{   3^{3/2} \pi
  M_D^{2/3} 
   \left(1+\frac{i \tau }{r}\right)^{2/3}},\\
I_{\mathrm{IR},-1}&=\int_{-\infty}^{\infty}\d
s\frac{\lambda^2(s-i)}{6\pi^2(s+i)(1+s^2)^{3/4}M_D} \Bigg(
\int_0^{\infty}\d u\Big(\frac{1}{u^{4/3}} - \frac{\theta(1-u)}{u^{4/3}\sqrt{1-u^{4/3}}}  \Big)  \Bigg) \non
&=\frac{\lambda^2}{6 \pi M_D}.
\end{align}
In total we have for large $\tilde{R}$:
\begin{align}
I=-\frac{\Gamma \left(\frac{2}{3}\right)\lambda^{2}\left|
   r\right| }{   3^{3/2} \pi
  M_D^{2/3} 
   \left(|r|+i\sgn (r) \tau\right)^{2/3}}  +  \frac{\lambda^2}{6\pi M_D}  +  \mathcal{O}(\lambda^{2}M_D^{-4/3}R^{-1/3}).
\end{align}
We see that the leading order term in $R$ is the same as was obtained from
the large $M_D$ approximation of the exponent. The first subleading
term is just a constant. This is good news because we already have the Fourier transform of this expression. This result is valid for length scales larger than
$1/M_D$ with a bounded error of the order $R^{-1/3}$. Defining this approximation as $G_{\mathrm{IR}}$, i.e.
\begin{align}
\label{eq:33}
G_\mathrm{IR}=G_0\exp\left(-\frac{\Gamma \left(\frac{2}{3}\right)\lambda^{2}\left|
   r\right| }{  3^{3/2} \pi
  M_D^{2/3} 
   \left(|r|+i\sgn (r) \tau\right)^{2/3}}  +  \frac{\lambda}{6 \pi M_D}\right)~,
\end{align}
the error of this approximation follows from:
\begin{align}
\Delta G_\mathrm{IR}=G-G_\mathrm{IR}=G_\mathrm{IR}\left(\exp\left( \mathcal{O}(\tilde{R}^{-1/3} )\right)-1\right).
\end{align}
Since the exponential in $G_\mathrm{IR}$ is bounded we have that $\Delta G_\mathrm{IR}=\mathcal{O}(r^{-4/3})$. After Fourier transforming this translates to an error of order $\mathcal{O}(k^{-2/3})$. %

\subsection{The exact fermion two-point function in momentum space:
  Numerical method}

Having understood the shortcomings of the naive large $M_D$ answer,
the way to derive the {\em exact} answer in real space, and the
correct IR approximation, we can now analyze the behavior of the
quantum critical metal at low energies. For this we need to transform
to frequency-momentum space. As our exact answer is in the form of an
infinite sum, this is not feasible analytically. We therefore resort
to a straightforward numerical Fourier analysis. 

To do so we first numerically determine the real space value of the
exact Green's functions. To do so accurately, several observations are
relevant
\begin{itemize}
\item The coefficients $f_n$ in the
infinite sum for $I(\tau,r)$ decay factorially in $n$ so once $n$ is
of order $\tilde{R}$, convergence is very rapid. 
\item
The hypergeometric functions for each $n$ are costly to compute with
high precision, but with the above choice of polar coordinates
the arguments of the hypergeometric functions are independent of $R$
and $M_D$. We therefore numerically evaluate the series
over a grid in $\tilde{R}$ and $\Phi$. We can then reuse the
hypergeometric function evaluations many times and greatly decrease
computing time.
\item 
The real space polar grid will be limited to a finite size. The IR expansion from Eq. \eqref{eq:33} can be used instead of 
the exact series for large enough $\tilde{R}$. To do so, we have to
ensure an overlapping regime of validity. It turns out that a rather
large value of $\tilde{R}$ is necessary to obtain 
numerical agreement between these two expansions, i.e. one
needs to evaluate a comparably large number of terms in the expansion. For the results
presented in this paper it has been necessary to compute coefficients
up to order 16 000 in $\tilde{R}$, for many different angles
$\Phi$. The function is bounded for large $\tau$ and $\tilde{R}$ but each term
grows quickly. This means that there are large cancellations between the terms that in the
end give us a small value. We therefore need to calculate these
coefficients to very high precision in evaluating the
polynomial. For
these high precision calculations, we have used the Gnu Multiprecision Library \cite{GMP}.
\item
On this polar grid we computed the exact answer for $\tilde{R}<\tilde{R}_0\approx1000$ and used
cubic interpolation for intermediate values.
For larger $\tilde{R}$ we use the
asymptotic expansion in Eq. \eqref{eq:33}. 
\end{itemize}

We then use a standard discrete numerical Fourier transform (DFT) to obtain
the momentum space two-point function from this numeric prescription
for $G(R,\Phi)$. Sampling $G(R,\Phi)$  at a finite number of discrete
points, the size of the sampling grid will introduce an IR cut-off at
the largest scales we sample and a UV-cut off set by the smallest
spacing between points. These errors in the final result can be
minimized by using the known asymptotic values analytically. Rather
than Fourier transforming $G(\tau,r)$ as a whole, we Fourier transform
$G_{\mathrm{diff}}(\tau,r)=G(\tau,r)-G_{\mathrm{IR}}(\tau,r)$
instead. Since both these functions approach the free propagator in
the UV, the Fourier transform of its difference will decay faster
for large $\omega$ and $k$. This greatly reduces the UV artefacts
inherent in a discrete Fourier transform. These two functions
{\em also} approach each other for large $\tau$ and $x$. In fact, with
the numerical method we use to approximate $f(\tilde{R},\Phi)$ described above, they
will be identical for $\tilde{R}_0<M_D\sqrt{\tau^2+r^2}$. This
means that we only need to sample the DFT within that area. With a DFT
we will always get some of the
UV tails of the function that gives rise to folding aliasing artefacts.
Now our function decays rapidly so one could do a DFT to very high frequencies and discard the high frequency part. This unfortunately takes up a lot of memory so we have gone with a more CPU intensive but memory friendly approach.
To address this we perform a convolution with a Gaussian
kernel, perform the DFT, keep the lowest 1/3 of the frequencies
and then divide by the Fourier transform of the kernel used. This gives us a good numeric value for
$G_{\mathrm{diff}}(\omega,k)$.
To this we add our analytic expression for $G_{\mathrm{IR}}(\omega,k)$.

\section{The physics of 2+1 quantum critical metals in double scaling limit} 
\label{sec:physics-2+1-quantum}

With the exact real space expression and the numerical momentum
space solution for the full non-perturbative fermion Green's
function, we can now discuss the physics of the elementary quantum
critical metal in the double scaling limit. Let us emphasize right
away that all our results are in Euclidean space. Although a Euclidean momentum space Green's function can be used to find a good Lorentzian continuation with a well-defined
and consistent spectral function, this function is not easily obtainable from
our numerical Euclidean result. We leave this for future work. The
Euclidean signature Green's function does not visually encode the spectrum
directly, but for very low energies/frequencies the Euclidean and the Lorentzian expressions are nearly
identical, and we can extract much of the IR physics already from the
Euclidean correlation function.

\begin{figure}[t!]
\centering{}
\hspace{-.2in}(A)
\includegraphics[width=0.37\textwidth]{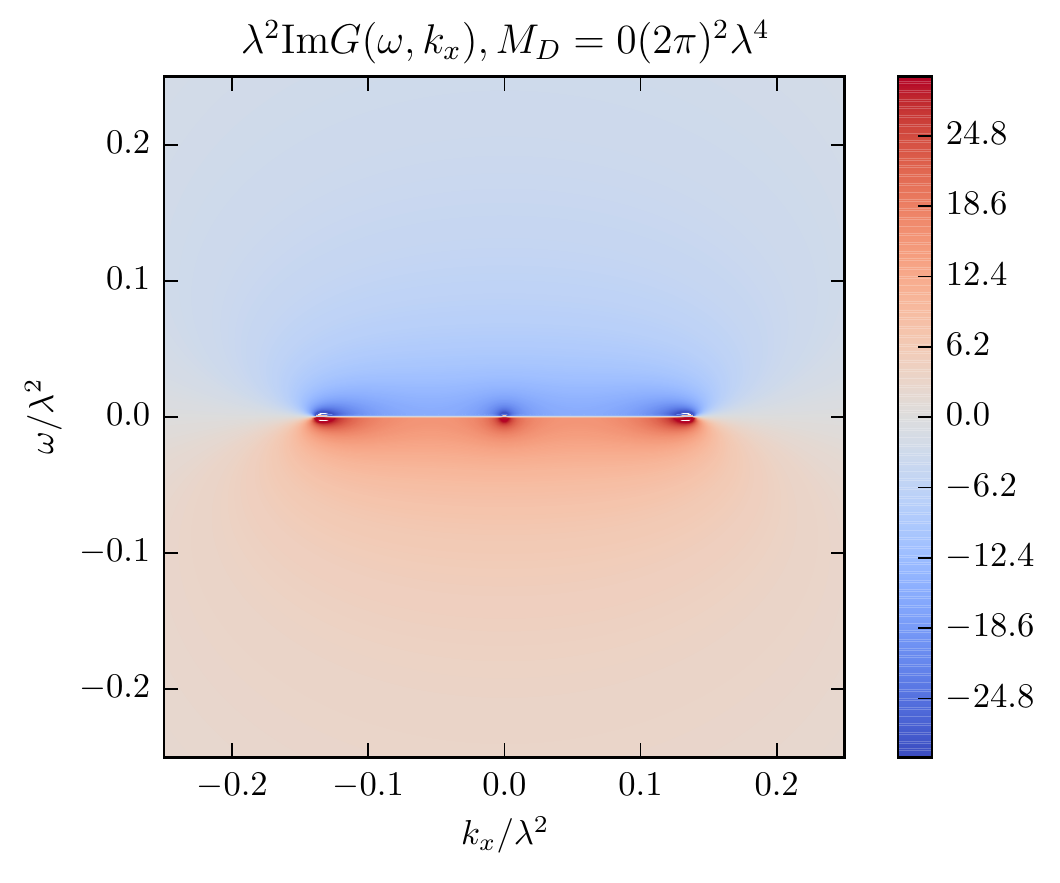}
\hspace{-.5in}
\includegraphics[width=0.37\textwidth]{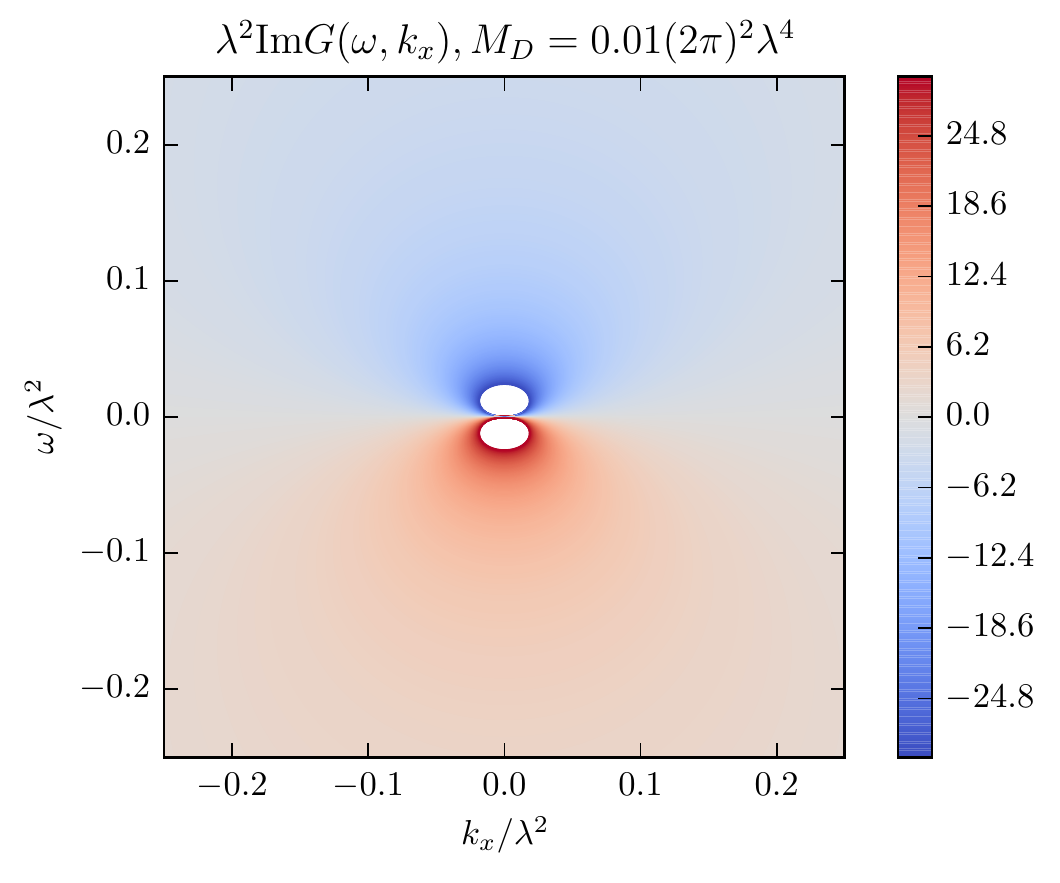}
\hspace{-.5in}
\includegraphics[width=0.37\textwidth]{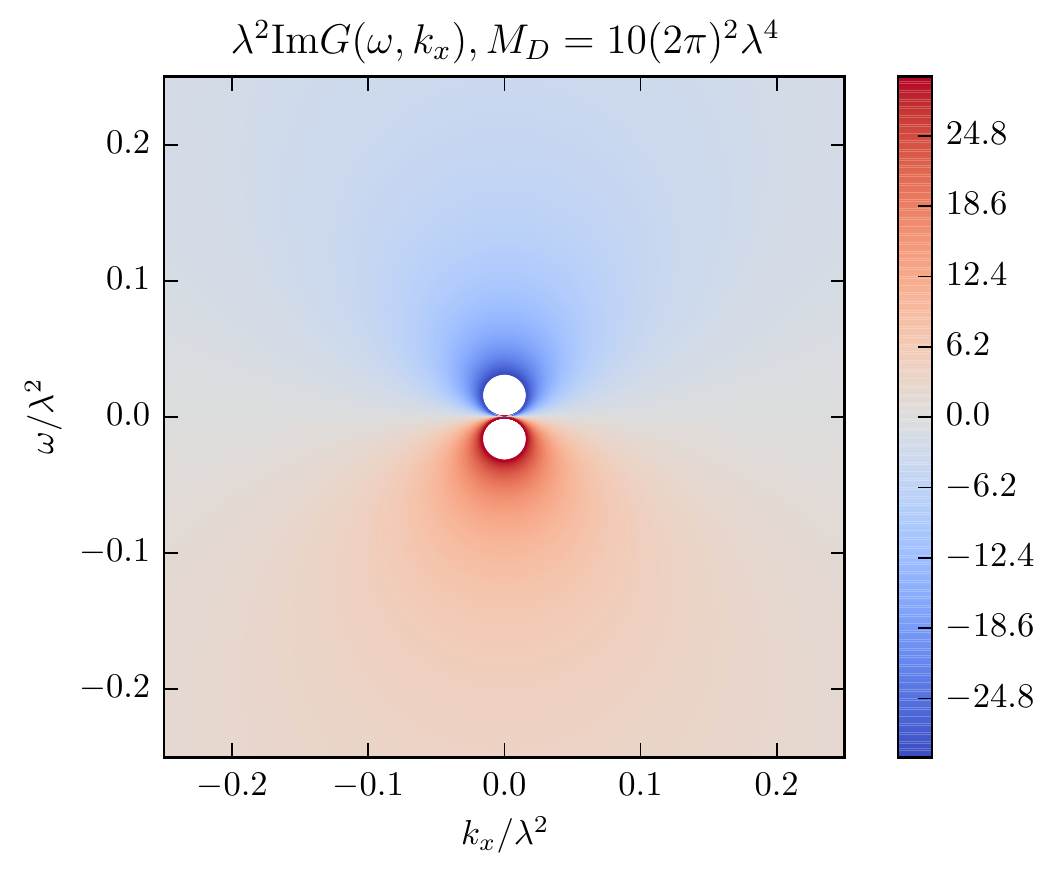}\\
\hspace{-.2in}(B)
\includegraphics[width=0.49\textwidth]{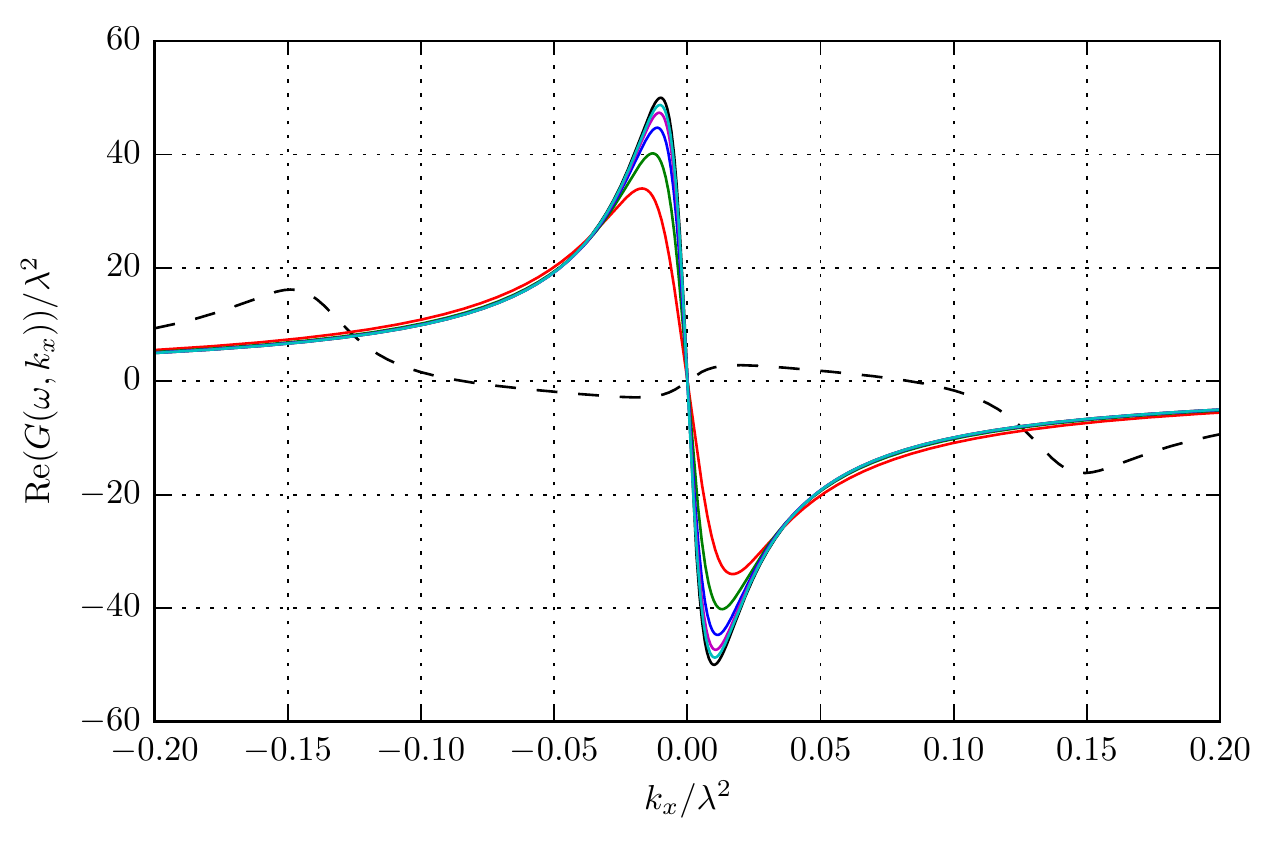}
\includegraphics[width=0.49\textwidth]{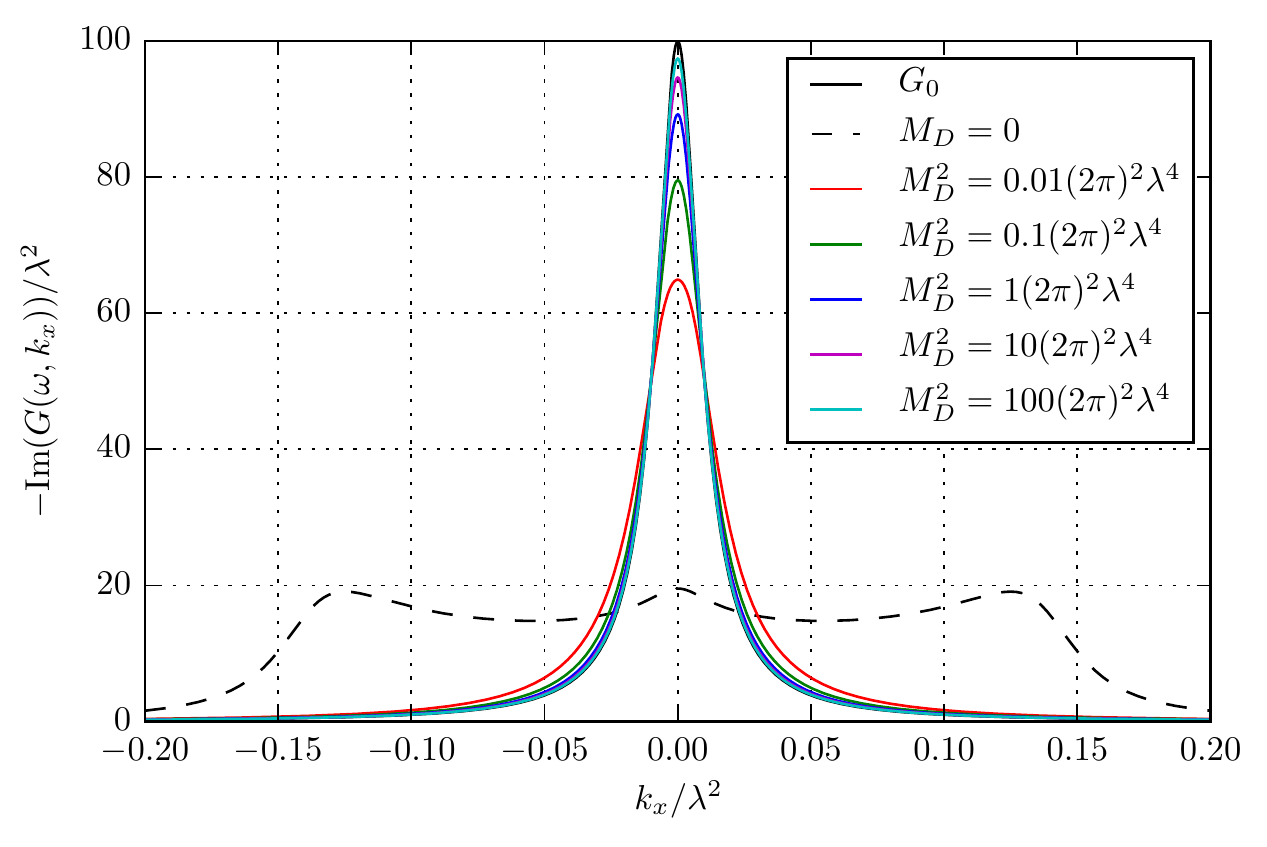}
\caption{(A) Density plots of the imaginary part of the exact (Euclidean)
  fermion Green's function $G(\omega,k_x)$ for various values of
  $M_D$. In the quenched limit $M_D=0$ the three Fermi surface
  singularities are visible. For any appreciable finite $M_D$ the Euclidean
  Green's function behaves as a single Fermi surface non-Fermi
  liquid. (B) Real and imaginary parts of $G(\omega,k_x)$ for very
  small $\omega=0.01\lambda^2$.
\label{fig:contour}}
\end{figure}

In Fig. \ref{fig:contour} we show density plots of the imaginary
part of $G(\omega,k_x)$ for different values of $M_D$ as well as
cross-sections at fixed low $\omega$. For the
formal limit
$M_D=0$ we detect three singularities near $\omega=0$ corresponding
with the three Fermi surfaces found in Lorentzian signature in our
earlier work \cite{quenched}. However, for any appreciable value of
the dimensionless ratio
$M_D/\lambda^2$ one only sees a single
singularity. As the plots for $G(\omega,k_x)$ at low frequency show,
its shape approaches that of the strongly Landau-damped $M_D\rar\infty$ result, Eq. \eqref{eq:GlargeNfAppEucResult1}, as one
increases $M_D/\lambda^2$, though for low $M_D$ it is still
distinguishably different. Recall that our results are derived in the
limit of large $k_F$ and therefore a realistic ($N_f \sim 1$) value for $M_D$ is $M_D\sim\lambda\sqrt{N_fk_F}\gg\lambda^2$.

This result is in contradistinction to what happens to the bosons.
When the bosons are not affected in the IR, i.e. the quenched limit, the fermions are greatly
affected by the boson: there is a topological Fermi surface transition
and the low-energy spectrum behaves as critical excitations
\cite{quenched}. However, once we increase $M_D$ to realistic
values, the  bosonic excitations are rapidly dominated in the IR by Landau
damping but we now see that this {\em reduces} the corrections to the
fermions. As $M_D$ is increased {for fixed $\omega$, $k_x$}, the deep IR fermion two-point function
approaches more and more that of the simple RPA result with self-energy $\Sigma\sim i\omega^{2/3}$.

A more careful analysis of the IR reveals that there are several distinct $\omega\rar0$ limits of the two-point function:
\begin{align*}
\lim_{\substack{\omega\rar0,\\k_x\text{ fixed}}} G_\mathrm{IR}(\omega,k_x)&\approx\begingroup\mathrm{e}^{\frac{\lambda^2}{6\pi M_D}}\endgroup\frac{1}{i\omega-k_x-\Sigma_{\mathrm{RPA}}}\\
 \lim_{\substack{\omega\rar0,\\k_x/\omega\text{ fixed}}}  G_\mathrm{IR}(\omega,k_x)  &\approx\begingroup\mathrm{e}^{\frac{\lambda^2}{6\pi M_D}}\endgroup\frac{1}{i\omega-k_x-\frac{4 \pi }{3 \sqrt{3}}\Sigma_{\mathrm{RPA}}}
\end{align*}
However in the case of $\omega^{2}\sim l_0k_x^3$, the full expression for $G_\mathrm{IR}$ is necessary to describe the low energy limit,
\begin{align}
\begin{split}
\lim_{\substack{\omega\rar0,\\l_0k_x^3\omega^{-2}\text{ fixed}}} G_{\mathrm{IR}}(\omega,k_x)=\mathrm{e}^{\frac{\lambda^2}{6\pi M_D}}\Bigg[
&\frac{1}{i\omega-k_x}  \cos \left(   \frac{\omega}{ l_0^{1/2}(\omega+ik_x)^{3/2}}      \right)\\
&+\frac{   6\sqrt{3}  i  \Gamma \left(\frac{1}{3}\right) \omega^{2/3} }{8 \pi l_0^{1/3}  (\omega+ik_x)^2} \,
   _1F_2\left(1;\frac{5}{6},\frac{4}{3};  - \frac{\omega^{2}}{4l_0(\omega+ik_x)^3}   \right) +\\
&+
       \frac{3  \sqrt{3}  i   \Gamma \left(-\frac{1}{3}\right)  \omega^{4/3}  }{8 \pi l_0^{2/3} (\omega+ik_x)^3}
 \,
   _1F_2\left(1;\frac{7}{6},\frac{5}{3}; - \frac{\omega^{2}}{4l_0(\omega+ik_x)^3}  \right)\Bigg],
   \end{split}
   \label{eq:GIRmom}
\end{align}

\begin{figure}
\centering{}
\hspace{-.2in}(A)
\includegraphics[width=0.49\textwidth]{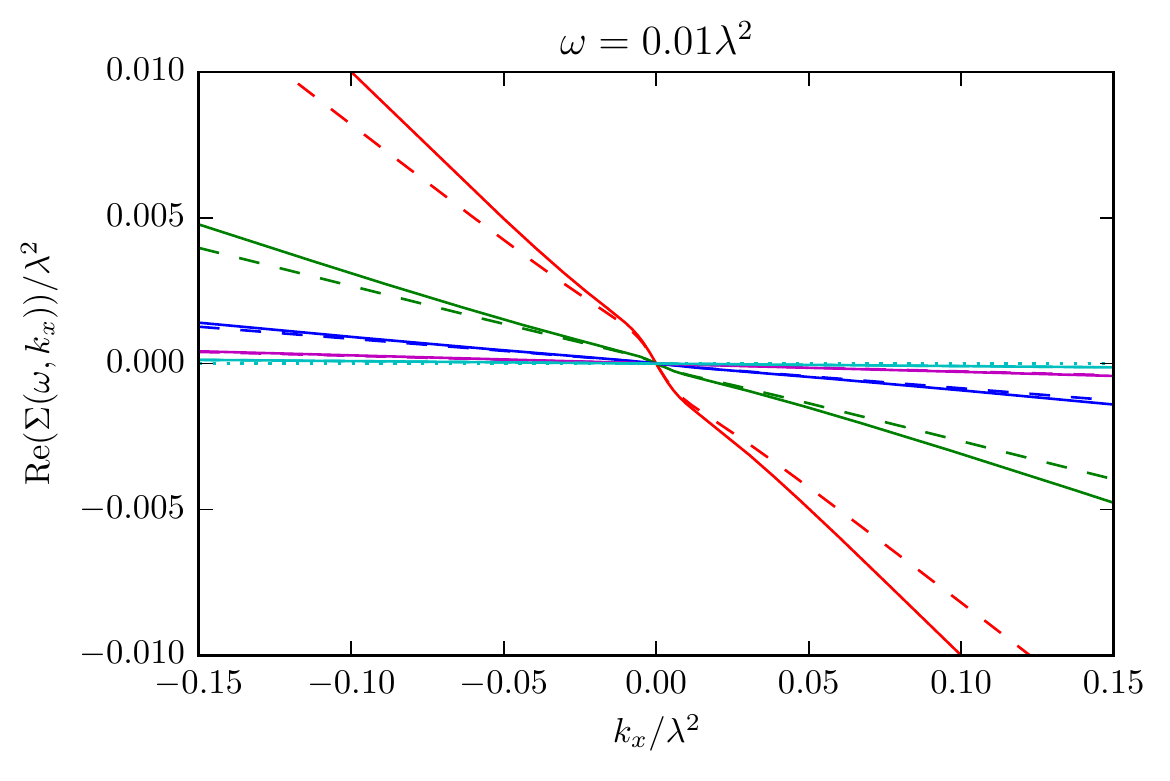}
\includegraphics[width=0.49\textwidth]{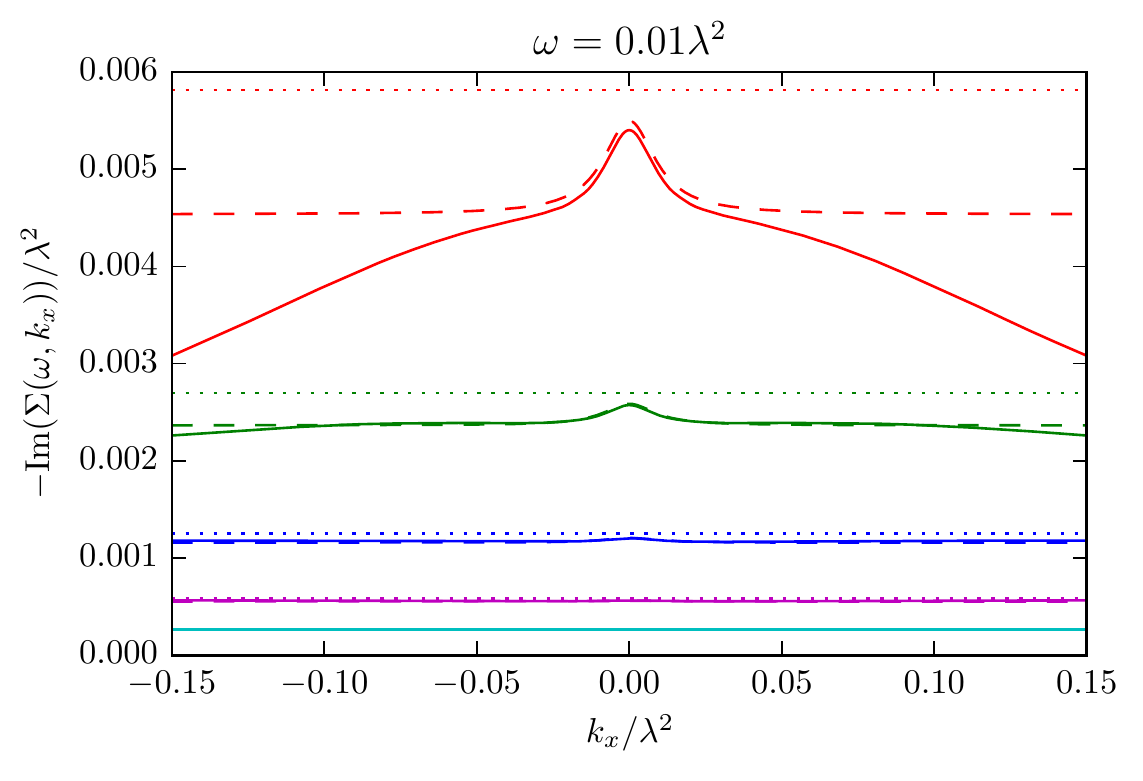}\\
\hspace{-.2in}(B)
\includegraphics[width=0.49\textwidth]{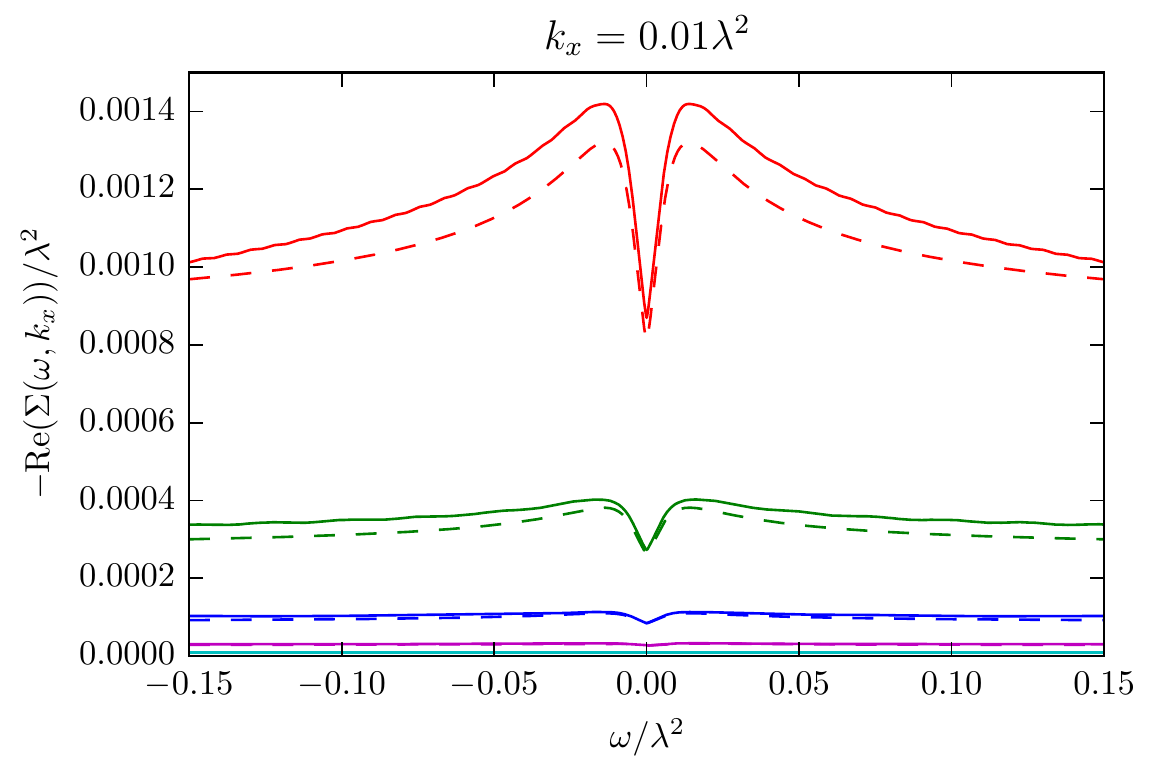}
\includegraphics[width=0.49\textwidth]{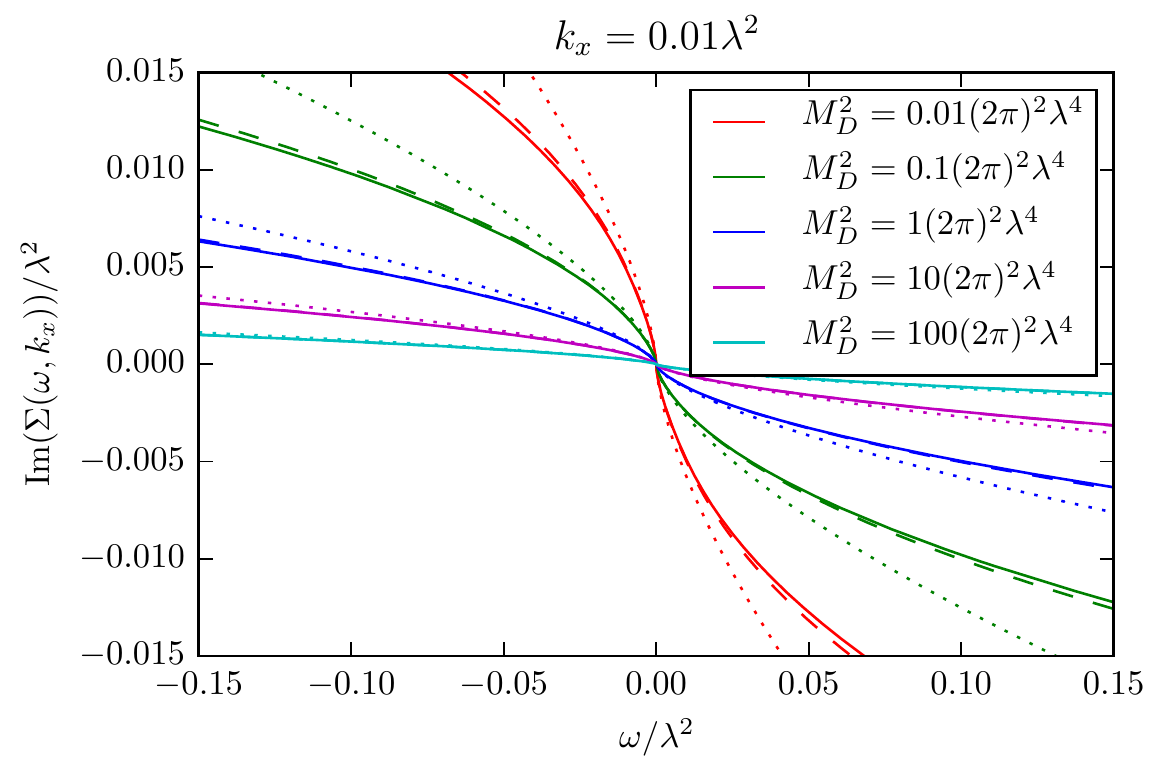}
\caption{Real and imaginary parts of the fermion self-energy for (A)
  $\omega=0.01\lambda^2$ and for (B) $k_x=0.01\lambda^2$. Dashed lines show the
  $G_{\mathrm{IR}}$-approximation; dotted lines show the RPA result.
 \label{fig:Sigma_vs_ks}}
\end{figure}

This existence of multiple limits shows in fact that the RPA result is never a good low energy (less than $M_D$) approximation for any value of $M_D/\lambda^2$.
We can illustrate this more clearly by studying the self-energy of the
fermion $\Sigma(\omega,k_x) = G(\omega,k_x)^{-1}-G_0(\omega,k_x)^{-1}$. 
It is shown in Fig. \ref{fig:Sigma_vs_ks} that the naive
large $M_D$ RPA result (dotted
lines) does agree for large $M_D$
at $\omega=0, k_x=0$ and the leading $\omega$ dependence of the
imaginary part is captured. The leading $k_x$
dependence is not captured by RPA. On the other hand, our improved
approximation for the low energy regime $G_{\mathrm{IR}}$ 
(dashed lines) captures these higher order terms in the low energy
expansion of $G(\omega,k_x)$ very well and also works for finite values of $M_D/\lambda^2$.

\begin{figure}
\centering{}
\includegraphics[width=0.49\textwidth]{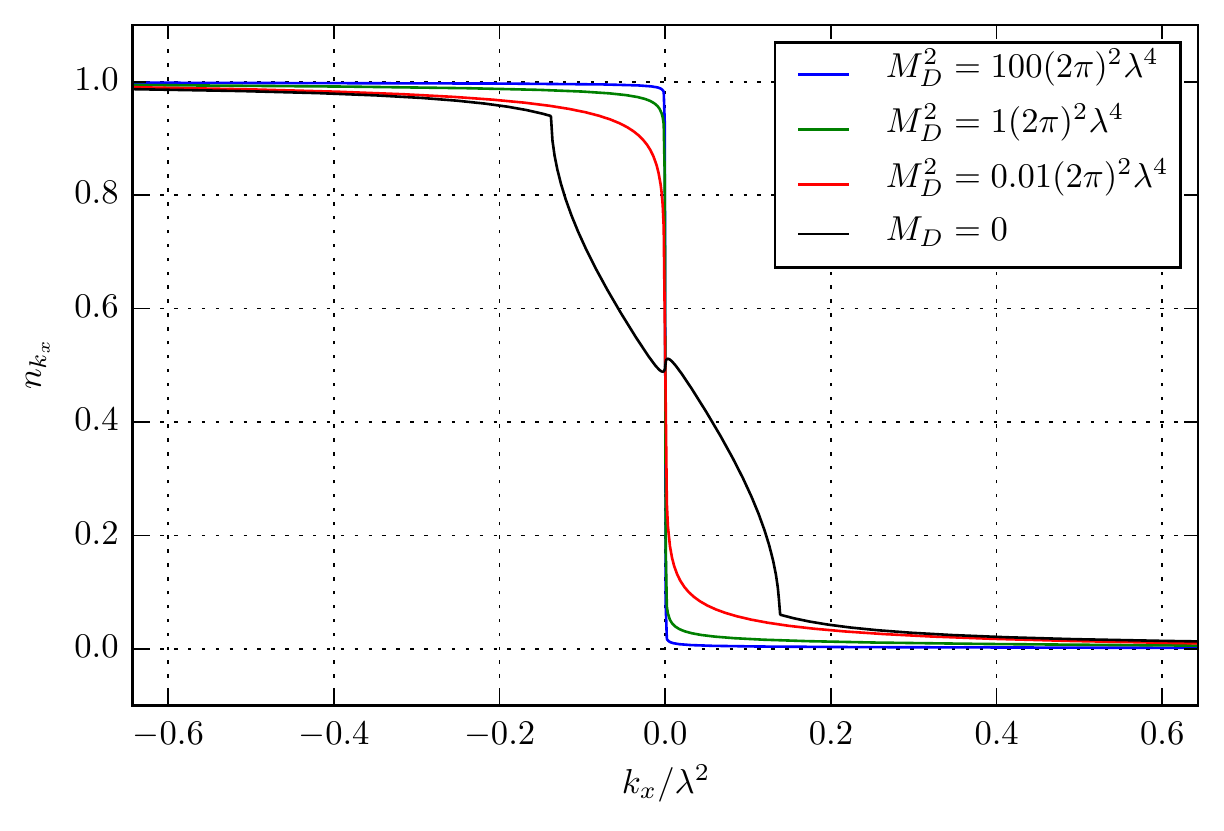}
\includegraphics[width=0.49\textwidth]{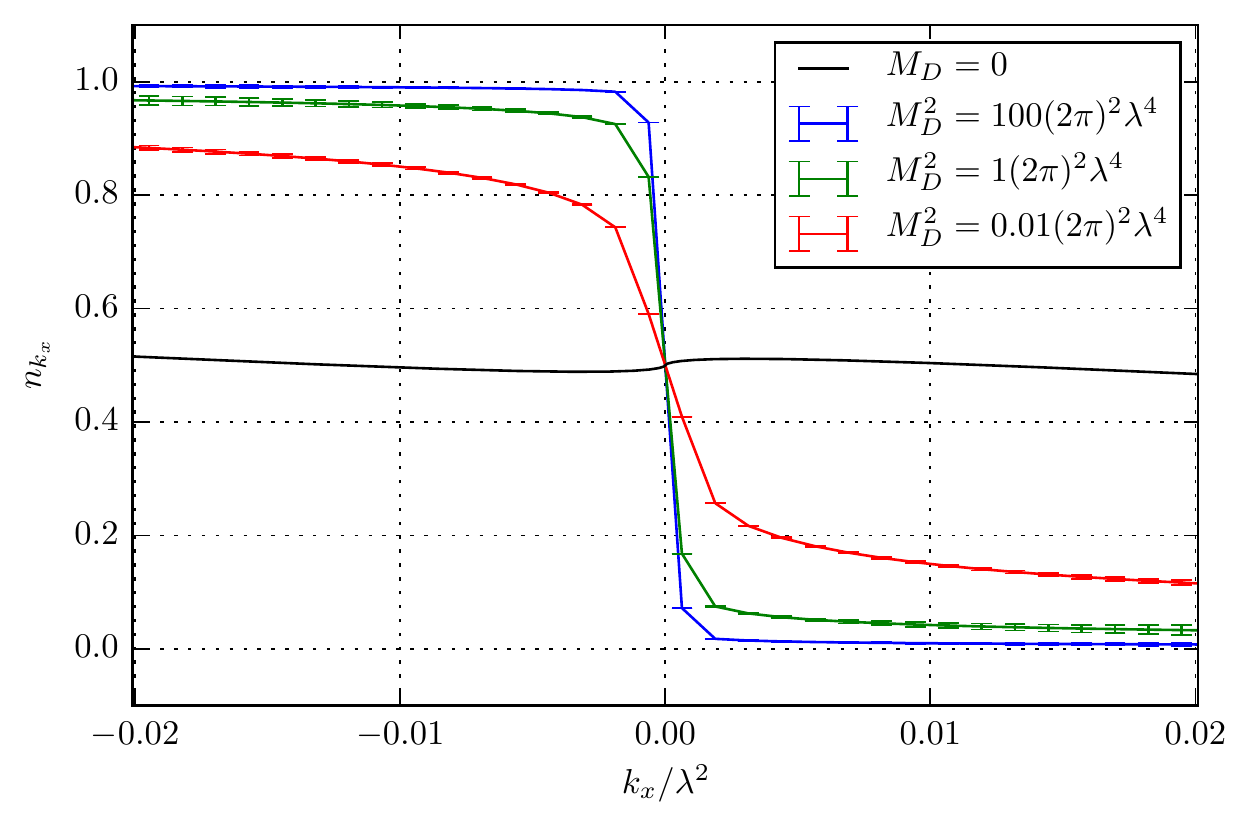}
\caption{Momentum distribution function. The two plots show the same function for different ranges. The error bars of the lower figure show an estimate of the error due to using the free Green's function close to the UV. The error-bars are exaggerated by a factor 100.\label{fig:n_vs_kx}}
\end{figure}

In all cases it is clear that the Fermion excitation, though qualitatively sharp, is not a Fermi liquid quasiparticle. We can calculate the occupation number and check whether it is consistent with the non-Fermi liquid nature of the Green's function. With a Fermi
liquid by definition is meant a spectrum with a discontinuity in the
zero-temperature  momentum distribution function
$n_{k}=\int_{-\infty}^0\dd \omega_R A(\omega_R,k_x)/2\pi$ with
$A(\omega_R,k_x)$ the spectral function. As the spectral function is
the imaginary part of the retarded Green's functions and the latter is
analytic in the upper half plane of $\omega_R$ we can move this
contour to Euclidean $\omega$ and use the fact that $G(\omega,k_x)$ approaches
$G_0(\omega,k_x)$ in the UV to calculate the momentum distribution function from
our Euclidean results. In detail
\begin{equation}
n_{k_x}=-\int_{-\infty}^0\dd \omega_R \im \frac{G_R(\omega_R,k_x)}{\pi}=\im\left [ \int_0^{\Lambda}\dd \omega i\frac{G(\omega,k_x)}{\pi}+\int_C\dd z i\frac{G(z,k_x)}{\pi}\right ]~;
\end{equation}
the first integral can be done with the numerics developed in the
preceding section. The contour $C$ goes from $\Lambda$ to $-\infty$
and for large enough $\Lambda$ this is in the UV and can well be
approximated by the free propagator. The resulting momentum
distribution $n(k)$ is shown in Fig. \ref{fig:n_vs_kx}. Within our
numerical resolution, these curves are continuous as opposed to a
Fermi liquid. This is of course expected; the continuity reflects the
absence of a clear pole in the IR expansions in the preceding
subsection.
Note also that as $M_D$ is lowered, the finite $M_D$ curves
approach the quenched result for $|k_x|>k_x^*$ where $k_x^*$ is the
point of the discontinuity of the derivative of the quenched
occupation number. At $k^{\ast}$ the (derivative) of the quenched
momentum distribution number does have a discontinuity (reflecting the
branch cut found in \cite{quenched}). 

{

\begin{figure}[t]
  \centering
  \includegraphics[width=0.8\textwidth]{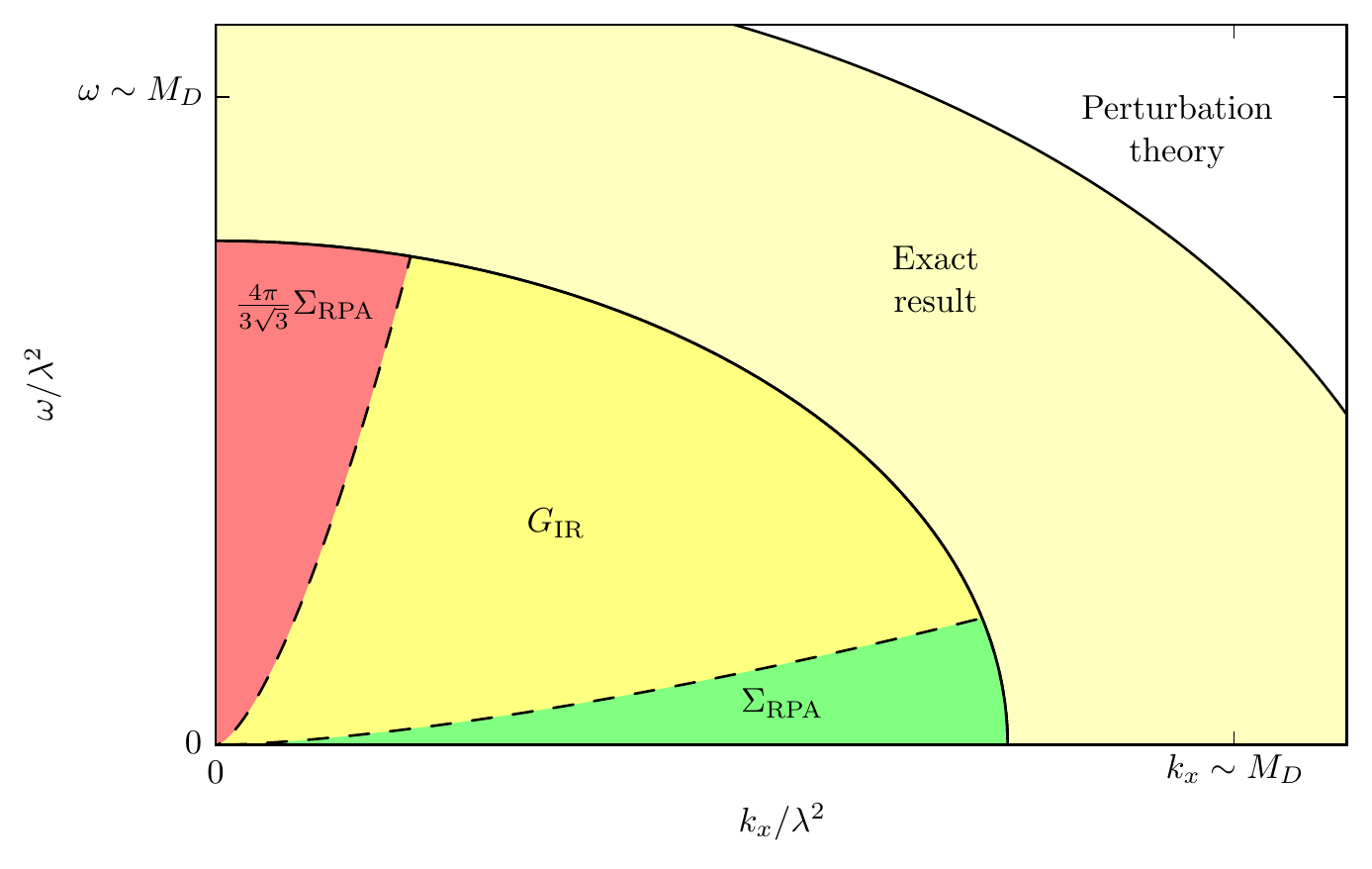}
  \caption{A sketch of the regimes of applicability of various approximations to the exact fermion Green's function of the elementary quantum critical metal in the double scaling limit. The $G_{\mathrm{IR}}$} approximation is applicable both in the deep yellow region and the red and green regions.
  \label{fig:sketch}
\end{figure}

\section{Conclusion}
\label{sec:conclusions}

We have presented a non-perturbative answer for the (Euclidean) fermion and boson two-point functions of the elementary quantum critical metal in the double limit of small $N_f$ and large $k_F$ with $N_fk_F$ fixed. {This limit was taken order by order in perturbation theory and additionally before performing bosonic momentum integrals.}%It would be enlightening to have our results in Lorentzian signature as well (as in the quenched limit); at the technical level this is an obvious next step.

Our exact results clarify how approximations that have been
made in the past hang together. The $N_f\rar0$, $k_F\rar \infty$ theory is
characterized by two energy scales $\lambda^2$ and $M_D$. For very large $\omega/\lambda^2$ and $\omega/M_D^2$ one can use perturbation theory in $\lambda$ to understand the theory. This is the perturbation around the UV-fixed point of a free fermion plus a free boson.
{
In the deep IR region, $\omega\ll M_D^2$, we have an analytic expression for the two-point function, $G_{\mathrm{IR}}$. This shows a $\omega^2\sim l_0k_x^3$ scaling.

In the intermediate regime, the full numerical results that we have presented are necessary.}

The quenched result we obtained earlier \cite{quenched} does not seem to have a useful regime of validity. As the momentum occupation number $n(k)$ indicates, its precise regime of applicability depends discontinuously on the
momentum $k/\lambda^2$. The discontinuity is surprising, but it can be
explained analytically as an order of limits ambiguity. Although it is hard to capture the deep IR region for very small $N_f$ in the
full numerics we {find from $G_{\mathrm{IR}}$ that} in the
$\omega$-$k$ plane there is a region where the limit $N_f \rar 0$ and
$\omega,k_x \rar 0$ do not commute. We show an indication of this in appendix~\ref{sec:cross-from-quench}. Thus for scales below $\lambda^2$, the quenched result is not useful. For scales above $\lambda^2$ with $M_D$ small, we can use perturbation theory so also in this regime the quenched result is not useful. For small $\omega$ and $k_x$, below $M_D$, the IR approximation we presented above gives a good description of the physics. We have presented a pictorial overview of how the various
approximations are related in Fig. \ref{fig:sketch}.

As argued before the spectrum of the elementary
quantum critical metal in the {ordered} double scaling limit obtained here is not the definitive one. {Taking this limit after the bosonic integrals and the perturbative sum may give a different result}. Another estimate for the true IR spectrum of the elementary quantum critical metal has been postulated before based on large $N_f$ approximations \cite{ChubukovNew, SungSik1,Mross:2010rd}, but our result at small $N_f$ gives a different IR.
Our non-perturbative answer follows from the insight that in the limit of large $k_F$ all fermion loops with
more than two external boson legs have cancellations upon symmetrization such that their scaling in $k_F$ is reduced. This higher loop cancellation has been previously put forward as a result of the strong forward scattering approximation. This approximation is engineered such that the Schwinger-Dyson equations combined with the fermion number Ward identity collapse to a closed set of equations. The solution to this closed set is the same dressed fermion correlator that we have presented. However the exact connection between the strong forward scattering approximation and the double scaling limit, is not clear. In the double scaling limit here, it is manifest that only the boson propagator is corrected and this in turn implies that the exact fermion Green's function in real space is an exponentially
dressed version of the free Green's function.

At the physics level, an obvious next step is therefore to explore the spectrum of the elementary quantum critical metal {including corrections in $N_f$ that are not proportional to $k_f$. Since this necessarily involves
  higher-point boson correlations,}  %
the role of self-interactions of the boson needs
to be considered. These are also relevant in the IR and may therefore
give rise to qualitatively very different physics than found here. We leave this for future work.

\acknowledgements
We are grateful to Andrey Chubukov, Sung-Sik Lee, Srinivas Raghu, Subir Sachdev, Jan Zaanen, and especially Vadim Cheianov for
discussions. 
This work was supported in part
by a VICI (KS) award of the
Netherlands Organization for Scientific Research (NWO), by the
Netherlands Organization for Scientific Research/Ministry of Science
and Education (NWO/OCW), by a Huygens Fellowship (BM), and by
the Foundation for Research into Fundamental Matter (FOM).

\appendix
\section{Multiloop cancellation}
\label{sec:mult-canc}

The robustness of the one-loop result in case of linear fermionic
dispersion was recognized before under the name of multiloop cancellation
\cite{Metzner97}. The technical result is that for a
theory with a simple Yukawa coupling and linear dispersion around a
Fermi surface, a symmetrized
fermion loop with more than two fermion lines vanishes. In our context
the linear dispersion is a consequence of the large $k_F$ limit. In
 other words all higher loop contributions to the polarization $\Pi$ should be
subleading in $1/k_F$. This was explicitly demonstrated at two loops
in  \cite{Kim2loop}.

We will give here a short derivation of this multiloop
cancellation in the limit $k_F\rar\infty$.
We may assume in this limit that the momentum transfer at any fermion
interaction is always much smaller than the size of the initial ($\vec{k}$) and
final momenta ($\vec{k}'$) which are of the order of the Fermi momentum, i.e. 
$|\vec{k}'-\vec{k}|\ll k_{F}$ with $|\vec{k}|, |\vec{k}'| \sim k_F$.
The free fermion Green's function then reflects a linear dispersion
\begin{align}
  \label{eq:18}
  G_0(\ome,k) = \frac{1}{i\ome - v k}
\end{align}
We now Fourier transform back to real space, as multiloop cancellation
is most easily shown in this basis.
The real space transform of the ``linear'' free fermion propagator
above is
\begin{align}
G_{0}\left(\tau,r\right)=-\frac{i}{2\pi}\frac{\sgn(v)}{r+iv\tau},
\end{align}
where as before $r$ is the conjugate variable to $k=|\vec{k}|-k_{F}$.
The essential step in the proof is that real space Green's function manifestly obeys the 
identity \cite{quenched}
\begin{align}
G_{0}\left(z_{1}\right)G_{0}\left(z_{2}\right)=G_{0}\left(z_{1}+z_{2}\right)\left(G_{0}\left(z_{1}\right)+G_{0}\left(z_{2}\right)\right)\label{eq:freeid}
\end{align}
with $z\equiv r+iv\tau$.
Consider then (the subpart of any correlation function/Feynman diagram
containing) a fermion loop with $n \geq 2$ vertices along the loop connected
to indistinguishable scalars
(i.e. no derivative interactions and all interactions are symmetrized). The corresponding algebraic
expression in a real space basis will then contain the expression
\begin{align}
F\left(z_{1},...,z_{n}\right)=\sum_{(i_{1},..,i_{n})\in S_{n}}G_{0}\left(z_{i_{1}}-z_{i_{2}}\right)G_{0}\left(z_{i_{2}}-z_{i_{3}}\right)...G_{0}\left(z_{i_{n-1}}-z_{i_{n}}\right)G_{0}\left(z_{i_{n}}-z_{i_{1}}\right),
\end{align}
where $S_{n}$ is the set of permutations of the numbers 1 through
$n$. Using the ``linear dispersion'' identity Eq. (\ref{eq:freeid}) 
and the shorthand notation
$G_{0}\left(z_{i_{1}}-z_{i_{2}}\right)=G_{12}$ we obtain
\begin{align}
\label{eq:20}
F\left(z_{1},...,z_{n}\right)=\sum_{(i_{1},..,i_{n})\in S_{n}}G_{12}G_{23}...G_{n-1,1}\left(G_{n-1,n}+G_{n,1}\right).
\end{align}
Next we cyclically permute the indices from $1$ to $n-1$: $1\rightarrow 2$,
$2\rightarrow3$, ..., $n-1\rightarrow1$ in the sum
\begin{align}
\sum_{(i_{1},..,i_{n})\in S_{n}}G_{12}G_{23}...G_{n-1,1}G_{n-1,n}=\sum_{(i_{1},..,i_{n})\in S_{n}}G_{23}G_{34}...G_{12}G_{1,n}.
\end{align}
This gives us
\begin{align}
F\left(z_{1},...,z_{n}\right)=\sum_{(i_{1},..,i_{n})\in S_{n}}G_{12}G_{23}...G_{n-1,1}\left(G_{1,n}+G_{n,1}\right).
\end{align}
Then since $G_{i,j}$ corresponds to a (spinless) fermionic Green's
function, it is antisymmetric $G_{i,j}=-G_{j,i}$, and we can conclude
that $F$ vanishes for $n\geq 3$. For $n=2$ %
 it is not possible to use the identity Eq. (\ref{eq:freeid}) since we would need to evaluate $G(0)$ which
is infinite.

\section{The Fourier transform of the fermion Green's function in the large $M_D$
  approximation}
\label{sec:ferm-greens-funct}

We will now show how to perform this Fourier transform. We need to calculate the following integral:
\begin{align}
\underset{M_D\rightarrow\infty}{G_f}(\omega,k)=\int\d\tau\d x\frac{\e^{i(\omega\tau-kx)}}{2\pi(ix-v\tau)}\exp\left(-\frac{\left| x\right| }{l_0^{1/3} \left(\left| x\right|+iv\sgn(x) \tau\right)^{2/3}}\right)
\label{cheatIntDef}
\end{align}
First we note that integrand is $\tau$-analytic in the region \linebreak$\{\tau\in\CC:\min(0,vx)<\im(\tau)<\max(0,vx)\}$. Since the integrand necessarily goes to 0 at $\infty$ we can thus shift the $\tau$ contour, $\tau\rightarrow\tau+ix/v$. We now have
\begin{align}
\underset{M_D\rightarrow\infty}{G_f}(\omega,k)=-\int\frac{\d\tau\d x}{2\pi v\tau}\exp\left(i\omega\tau -x\left( i k+\omega/v +\frac{\sgn(x) }{l_0^{1/3} \left(iv\sgn(x) \tau\right)^{2/3}}\right)\right)
\end{align}
We have allowed ourselves to choose the order of integration, shift the contour, and then change the order. We see that the $x$ integral now is trivial but only converges for
\begin{align}
-\frac{v^{1/3}}{2l_0^{1/3} |\tau|^{2/3}}<\re(\omega) <\frac{v^{1/3}}{2l_0^{1/3} |\tau|^{2/3}}
\end{align}
This is fine since we know that the final result is $\omega$-analytic in both the right and left open half-planes. As long as we can obtain an answer valid within open subsets of both of these sets we can analytically continue the found solution to the whole half planes. We thus proceed assuming $\re(\omega)$ is in this range. One can further use symmetries of our expression to relate the left and right $\omega$-half-planes, so to simplify matters, from now on we additionally assume $\omega$ to be positive. Let us now consider a negative $x$, we then see that the $\tau$ integral can be closed in the upper half plane and since it is holomorphic there the result will be 0. We can thus limit the $x$ integrals to $\RR^+$. We then have
\begin{align}
\underset{`N_fk_F\rightarrow\infty}{G_f}(\omega,k)=\int\d\tau\frac{\e^{i\omega\tau}}{2\pi v\tau}\frac{-1}{ a +\frac{1 }{l_0^{1/3} \left(iv\tau\right)^{2/3}}}
\end{align}
where $a=i k+\omega/v$. The integrand has a pole at $\left(i\tau\right)^{2/3} = -1/(al_0^{1/3}v^{2/3})$. Now break the integral in positive and negative $\tau$ and write it as
\begin{align}
\underset{M_D\rightarrow\infty}{G_f}(\omega,k)=\frac{h((-i)^{2/3}u_0^*)^*-h((-i)^{2/3}u_0)}{2\pi v a}
\end{align}
where
\begin{align}
\begin{split}
h(u)&=\int_0^\infty \d\tau\frac{\e^{i\tau}}{\tau+u\tau^{1/3}},\\
u_0&=\frac{\omega^{2/3}}{al_0^{1/3}v^{2/3}}.
\end{split}
\end{align}
Usng Morera's theorem we can prove that $h$ is holomorphic on $\CC/\RR^-$. Consider any closed curve $C$ in $\CC/\RR^-$. We need to show that
\begin{align}
\int_C\d u h(u)=0
\end{align}
We do this by rotating the contour slightly counter clockwise. For any curve $C$ there is clearly a small $\epsilon>0$ such that we will still not hit the pole $\tau+u\tau^{1/3}=0$.
\begin{align}
\int_C\d u \int_0^{(1+i\epsilon)\infty} \d\tau\frac{\e^{i\tau}}{\tau+u\tau^{1/3}},
\end{align}
The piece at $\infty$ converges without the denominator and thus goes to 0. Since the integral now converges absolutely we can use Fubini's theorem to change the orders of integration
\begin{align}
 \int_0^{(1+i\epsilon)\infty} \d\tau \int_C\d u\frac{\e^{i\tau}}{\tau+u\tau^{1/3}}=0
\end{align}
And since also the integrand is holomorphic on a connected open set containing $C$ the proof is finished.

The function $h$ can for $0<\mathrm{arg}(u)<2\pi/3$ be expressed as a Meijer G-function:
\begin{align}
h(u)=
\frac{3}{8 \pi ^{5/2}} G_{3,5}^{5,3}\left(
\begin{array}{c}
 0,\frac{1}{3},\frac{2}{3} \\
 0,0,\frac{1}{3},\frac{1}{2},\frac{2}{3} \\
\end{array}
\Bigg|-\frac{u^3}{4}\right)
\end{align}
The G-function has a branch cut at $\RR^-$ and because of that we can not easily write $h(u)$ in terms of it since the argument $u$ appears cubed in the argument of the G-function and we know that $h$ is holomorphic on all of $\CC/\RR-$. We can however express $h$ as an analytic continuation of the G-function past this branch cut, onto the further sheets of its Riemann surface.
\begin{figure}
\centering{}
\includegraphics{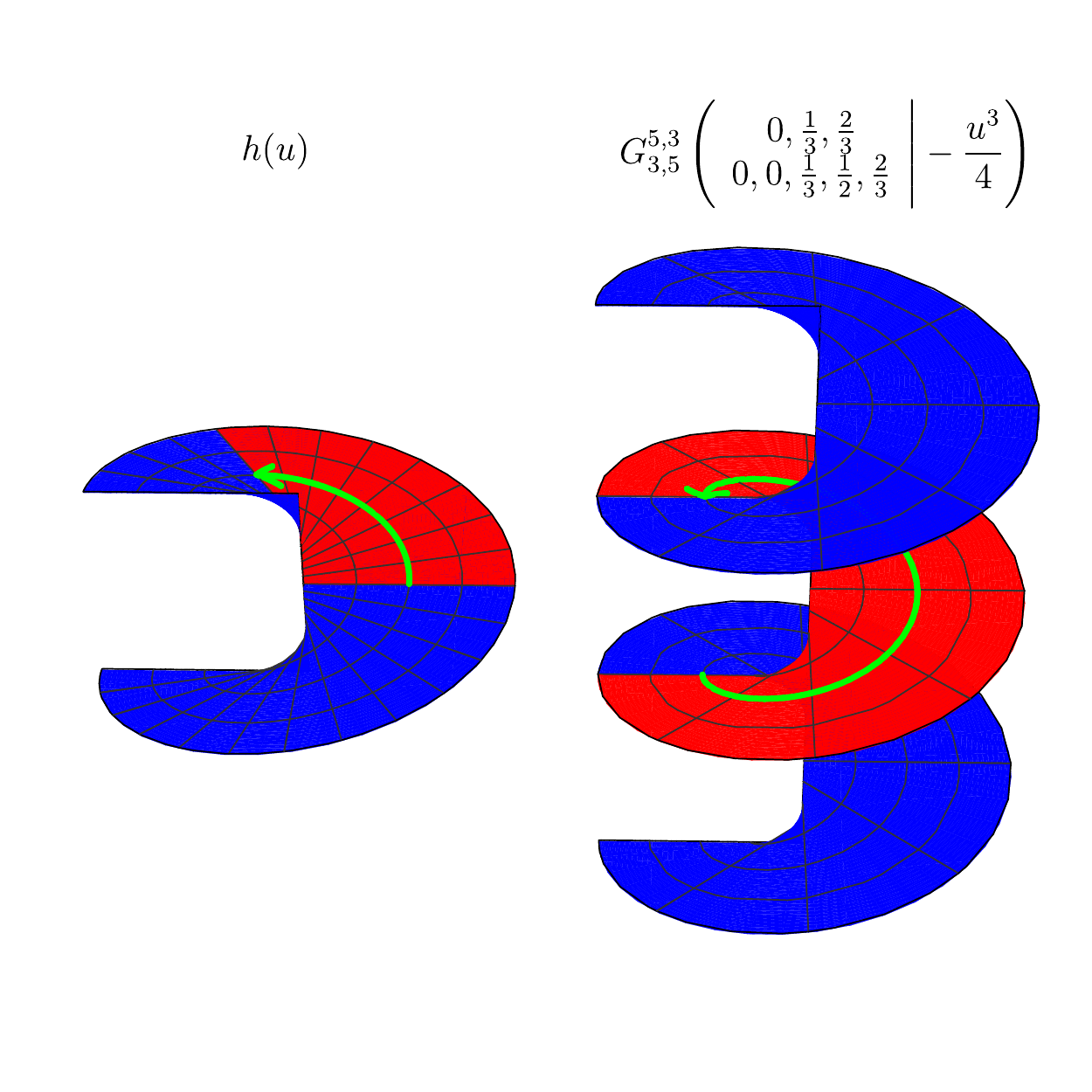}
\caption{The left image shows part of the Riemann surface of the function $h$ and the right image shows part of the Riemann surface of the G-function. The part in red of the left image can be expressed as the G-function evaluated on its first sheet, the red part of the right image. The green arrow shows how points are mapped from one Riemann surface to the other by the mapping $u\mapsto-u^3/4$\label{fig:riemannSurfaces}}
\end{figure}
We will write the G-function as a function with two real arguments, first the absolute value and second the phase of its otherwise complex argument. We will allow the phase to be any real number and when outside the range $[-\pi,\pi]$, we let the function be defined by its analytical continuation to the corresponding sheet. We hereafter omit the constant parameters of the G-function. We then have:
\begin{align}
\begin{split}
\underset{M_D\rightarrow\infty}{G_f}(\omega,k)&=
3\frac{
 G_{3,5}^{5,3}\big(
\frac{|u_0|^3}{4}  ,-2\pi - 3\arg(u_0)  \big)^*
-
 G_{3,5}^{5,3}\big( \frac{|u_0|^3}{4} , -2\pi +3\arg(u_0)\big)
}{16 \pi ^{7/2} v a}\\
&=
3\frac{
 G_{3,5}^{5,3}\big(
\frac{|u_0|^3}{4}  , 3\arg(u_0) +2\pi  \big)
-
 G_{3,5}^{5,3}\big( \frac{|u_0|^3}{4} ,3\arg(u_0) -2\pi\big)
}{16 \pi ^{7/2} v a}
\end{split}
\end{align}
In the last step we used the fact that the G-function commutes with complex conjugation. We see that the Green's function is given by a certain monodromy of the G-function. It is given by the difference in its value starting at a point $u_0^3/4$ on the sheet above the standard one and then analytically continuing clockwise around the origin twice to the sheet below the standard one and there return to $u_0^3/4$. Since we only need this difference, we might expect this to be a, in some sense, simpler function as would happen for e.g. monodromies of the logarithm. To see how to simplify this we look at the definition of the G-function. It is defined as an integral along $L$:
\begin{align}
G_{p,q}^{\,m,n} \!\left( \left. \begin{matrix} a_1, \dots, a_p \\ b_1, \dots, b_q \end{matrix} \; \right| \, z \right) = \frac{1}{2 \pi i} \int_L \frac{\prod_{j=1}^m \Gamma(b_j - s) \prod_{j=1}^n \Gamma(1 - a_j +s)} {\prod_{j=m+1}^q \Gamma(1 - b_j + s) \prod_{j=n+1}^p \Gamma(a_j - s)} \,z^s \,ds,
\end{align}
There are a few different options for $L$ and which one to use depends on the arguments. In our case $L$ starts and ends at $+\infty$ and circles all the poles of $\Gamma(b_i-s)$ in the negative direction. Using the residue theorem we can recast the integral to a series. We have double poles at all negative integers and some simple poles in between. The calculation to figure out the residues of all these single and double poles is a bit too technical to present here but in the end the series can be written as
\begin{align}
\begin{split}
G_{3,5}^{5,3}\left(z\right)=\sum_{n=0}^\infty\Big(a_n z^{n}+b_n z^{n}\log(z)+c_n z^{n+1/3}+d_n z^{n+1/2}+e_n z^{n+2/3}\Big)
\end{split}
\end{align}
Now we perform the monodromy term by term and a lot of these terms cancel out.
\begin{align}
\begin{split}
 &G_{3,5}^{5,3}\big(
|z|  , \arg(z) +2\pi  \big)
-
 G_{3,5}^{5,3}\big(|z|  , \arg(z)  -2\pi\big) =\\
 &\sum_{n=0}^\infty\Big(4\pi ib_n z^{n}+i \sqrt{3} c_n z^{n+1/3} -i \sqrt{3} e_n z^{n+2/3}\Big)
\end{split}
\end{align}
The coefficients $a_i$ contain both the harmonic numbers and the polygamma function whereas the other coefficients are just simple products of gamma functions. This simplification now lets us sum this series to a couple of generalized hypergeometric functions. Inserting the expressions for $a$ and $u_0$ we have
\begin{align}
\begin{split}
\underset{M_D\rightarrow\infty}{G_f}(\omega,k)=
&\frac{1}{i\omega-k v}  \cos \left(   \frac{\omega}{v l_0^{1/2}(\omega/v+ik)^{3/2}}      \right)\\
&+\frac{   6\sqrt{3}  i  \Gamma \left(\frac{1}{3}\right) \omega^{2/3} }{8 \pi l_0^{1/3}v^{5/3}  (\omega/v+ik)^2} \,
   _1F_2\left(1;\frac{5}{6},\frac{4}{3};  - \frac{\omega^{2}}{4l_0v^2(\omega/v+ik)^3}   \right) +\\
&+
       \frac{3  \sqrt{3}  i   \Gamma \left(-\frac{1}{3}\right)  \omega^{4/3}  }{8 \pi l_0^{2/3}v^{7/3} (\omega/v+ik)^3}
 \,
   _1F_2\left(1;\frac{7}{6},\frac{5}{3}; - \frac{\omega^{2}}{4l_0v^2(\omega/v+ik)^3}  \right) .
   \end{split}
   \label{eq:GlargeNfAppEucResult}
\end{align}%
Note that this expression is $\omega$-holomorphic for $\omega$ in the right half plane so our previous assumptions on $\omega$ can be relaxed as long as $\omega$ is in the right half plane. We note from expression \eqref{cheatIntDef} that if we change sign on both $\omega$ and $k$ and do the changes of variables $\tau\rightarrow-\tau$ and $x\rightarrow-x$ we end up with the same integral up to an overall minus sign. We can thus get the left half plane result using the relation
\begin{align}
\underset{M_D\rightarrow\infty}{G_f}(-\omega,k)=-\underset{M_D\rightarrow\infty}{G_f}(\omega,-k).
\end{align}
As mentioned in the main text, this expression has been compared with
numerics to verify that we have not made any mistakes. See
Fig. \ref{fig:benchmark}. We have also done the two integrals for the
Fourier transform in the opposite order, first obtaining a different
Meijer G-function then using the G-function convolution theorem to do
the second integral. In the end one obtains the same monodromy of the G-function as above. {This expression can also be found in Appendix A.2 of \cite{ChubukovNew}, however it was not entirely clear from the phrasing of the last paragraph that this function is actually the exact Fourier transform, Eq. \eqref{cheatIntDef}.}

\section{The discontinuous transition from the quenched to the
  Landau-damped regime}
\label{sec:cross-from-quench}

We show here why including Landau damping (finite $M_D$) physics
starting from the quenched $N_{f}\rightarrow0$ result, is
discontinuous in the IR. To do so we
calculate the Green's function by imposing the $N_f\rar 0$ limit from
the begining. We need to evaluate the Fourier transform integral:

\begin{align}
G_{L}\left(\omega,k\right)=\int_{-\infty}^{\infty}dx\int_{-L}^{L}d\tau G_{0}\exp\left(I_{0}+i\omega\cdot\tau-ik\cdot x\right),\label{eq:GL}
\end{align}
where as before the free propagator is

\begin{align}
G_{0}(\tau,x)=-\frac{i}{2\pi}\frac{1}{x+i\cdot\tau}
\end{align}
and the $N_{f}\rightarrow0$ limit of the exponent of the real space
Green's function: 
\begin{align}
I_{0}=\frac{(\tau+i\cdot x)^{2}}{12\pi\sqrt{\tau^{2}+x^{2}}}.
\end{align}

Note however that the $\tau$ integral is divergent in this limit.
Therefore in (\ref{eq:GL}) we have introduced a cutoff $L$ in this
direction. By looking at the full expansion of $I$ we can see that
the natural value of $L$ is of the order $1/M_D$. For larger
values of $\tau$ the asymptotic expansion describes $I$ better.
We expect that for large enough momenta and frequencies the asymptotic
region does not contribute to the Fourier transform and therefore
the cutoff can be removed. This naive epxectation however is only
partly true. We will shortly see that the region in the $\omega-k$
plane where the cutoff can be removed is more complicated and asymmetric
in terms of the momenta and frequency.

Let us turn now to the evaluation of (\ref{eq:GL}). After making
the coordinate change $\tau\rightarrow\tau$, $x\rightarrow u\cdot\tau$
one of the integrals ($\tau$) can be evaluated analytically:
\begin{align}
G_{L}\left(\omega,k\right)=\int_{-\infty}^{\infty}du\left(I_{1}(u)+I_{2}(u)+I_{3}(u)\right),
\end{align}
where
\begin{align}
I_{1}(u)=\frac{6\exp\left(ikLu\sqrt{u^{2}+1}-iL\cdot\omega\cdot\sqrt{u^{2}+1}-\frac{L(u-i)^{2}}{12\pi}\right)}{12\pi(u+i)(ku-\omega)+i\sqrt{u^{2}+1}u+\sqrt{u^{2}+1}},\label{eq:I1}
\end{align}
\begin{align}
I_{2}(u)=\frac{6i\exp\left(-ikLu\sqrt{u^{2}+1}+iL\cdot\omega\cdot\sqrt{u^{2}+1}-\frac{L(u-i)^{2}}{12\pi}\right)}{12i\pi k(u+i)u+\sqrt{u^{2}+1}u-i\sqrt{u^{2}+1}+12\pi(1-iu)\omega},\label{eq:I2}
\end{align}
\begin{align}
I_{3}(u)=-\frac{144\pi(ku-\omega)}{u\left(-3+u\left(144\pi^{2}k^{2}(u+i)+u-3i\right)\right)-288\pi^{2}ku(u+i)\omega+144\pi^{2}(u+i)\omega^{2}+i}.\label{eq:I3}
\end{align}
By numerically performing the single integral $u$ we can obtain $G_{M_D\rightarrow0}$.
Since it is easier than evaluating the Fourier transform of the true
real-space version of the Green's function it is worth understanding
how the $L\rightarrow\infty$ (which is equivalent to $M_D\rightarrow0$)
works. The result is depicted on Fig. \ref{fig:convergence}. In the
shaded region (which corresponds to small $k$) the limit is not well
defined while outside of this region the limit is equal to the quenched
result. The numerics shows that for zero frequency, the edge of this
region is at $k^{*}$, where the Green's function is singular.

\begin{figure}
\begin{centering}
\includegraphics{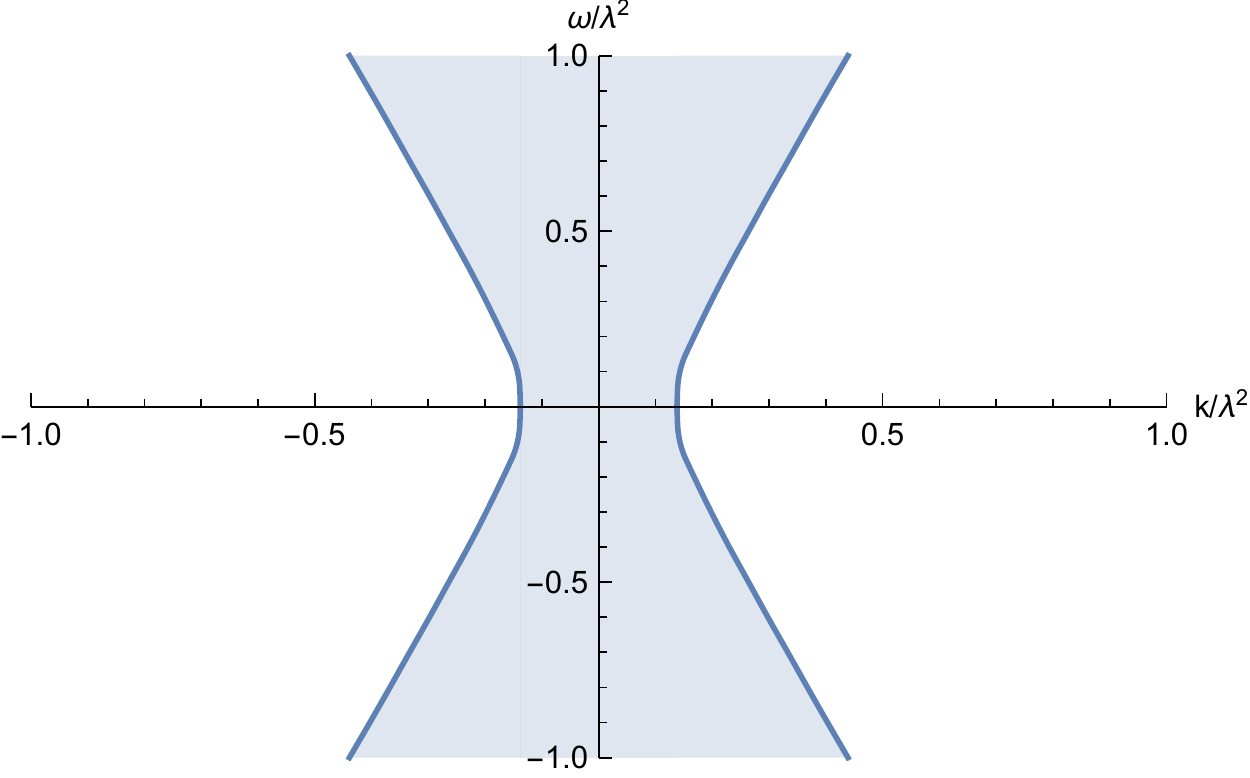}
\par\end{centering}

\caption{The region of convergence. In the shaded area the $L\rightarrow\infty$
($N_{f}\rightarrow0$) limit is not convergent while outside this
area $\lim_{L\rightarrow\infty}G_{L}=G_{quenched}$.\label{fig:convergence}}
\end{figure}

We can qualitatively determine the line separating the convergent
and divergent region. For this we assume that when $L$ is large one
can expand the exponent in $u$
\begin{align}
I_{1}(u)\sim\exp\left[u^{2}\left(-\frac{1}{12\pi}-\frac{1}{2}i\omega\right)L+\frac{iu(6\pi k+1)}{6\pi}L-iL\omega+\frac{L}{12\pi}+{\cal O}\left(L\dot{u^{3}}\right)\right].
\end{align}
We see that because of the term $-L\cdot u^{2}/(12\pi)$, the integrand
is non-zero only in a narrow region around $u=0$. For the same reason
we approximate the denominator of (\ref{eq:I1}) by replace $u$ by
zero there. With these simplifications we arrive to a gaussian integral
which can be evaluated analytically:
\begin{align}
\int_{-\infty}^{\infty}I_{1}(u)du\approx\frac{12i\sqrt{3}\pi\exp\left(\frac{iL\left(6\pi k^{2}+2k+\omega(-12\pi\omega+i)\right)}{12\pi\omega-2i}\right)}{(12\pi\omega+i)\sqrt{L(1+6i\pi\omega)}}.\label{eq:approxI1}
\end{align}
The real part of (\ref{eq:approxI1}) is
\begin{align}
L\frac{3\left(\omega^{2}-k^{2}\right)\pi-k}{36\pi^{2}\omega^{2}+1}.
\end{align}
It is clear that if this value is positive (i.e. $3\left(\omega^{2}-k^{2}\right)\pi-k>0$)
than the $L\rightarrow\infty$ limit is divergent. Looking at the
numerical result in Fig. \ref{fig:convergence} we indeed see that
the boundary of the shaded region is indeed a hyperbola. Note, however,
that the exact location of this hyperbola obtained from expanding
the exponent is slightly off.

It is interesting to note that if we are in the divergent region $G_{L}$
is not convergent for large $L$ but for some intermediate values
it can still get close to the quenched result $G_{quenched}$. To
quantify this let us introduce a relative ``error'' function
\begin{align}
error(L)=\left|\frac{G_{L}-G_{quenched}}{G_{quenched}}\right|.
\end{align}
In Fig. \ref{fig:error} we show the behavior of this function for
various points in the $\omega-k$ plane. For a point which is outside
the shaded region the error approaches zero when $L$ is large. For
$\omega=1$, $k=0$ which is inside the divergent region the error
is oscillating but there is an interval of $L$ where the amplitude
of this oscillation has a minimum. If the frequency is small
($\omega=0.1$), the amplitude is larger.%

\begin{figure}
\begin{centering}
\includegraphics[width=0.33\textwidth]{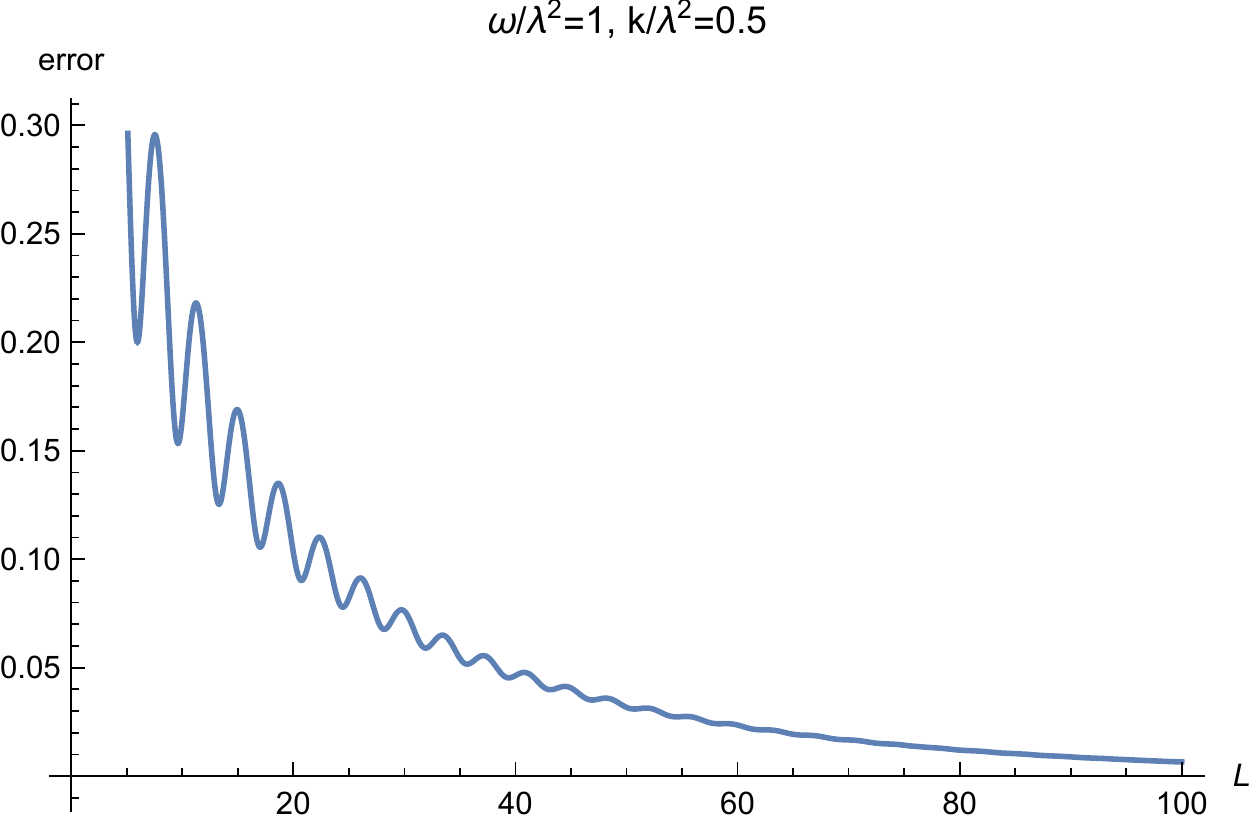}\includegraphics[width=0.33\textwidth]{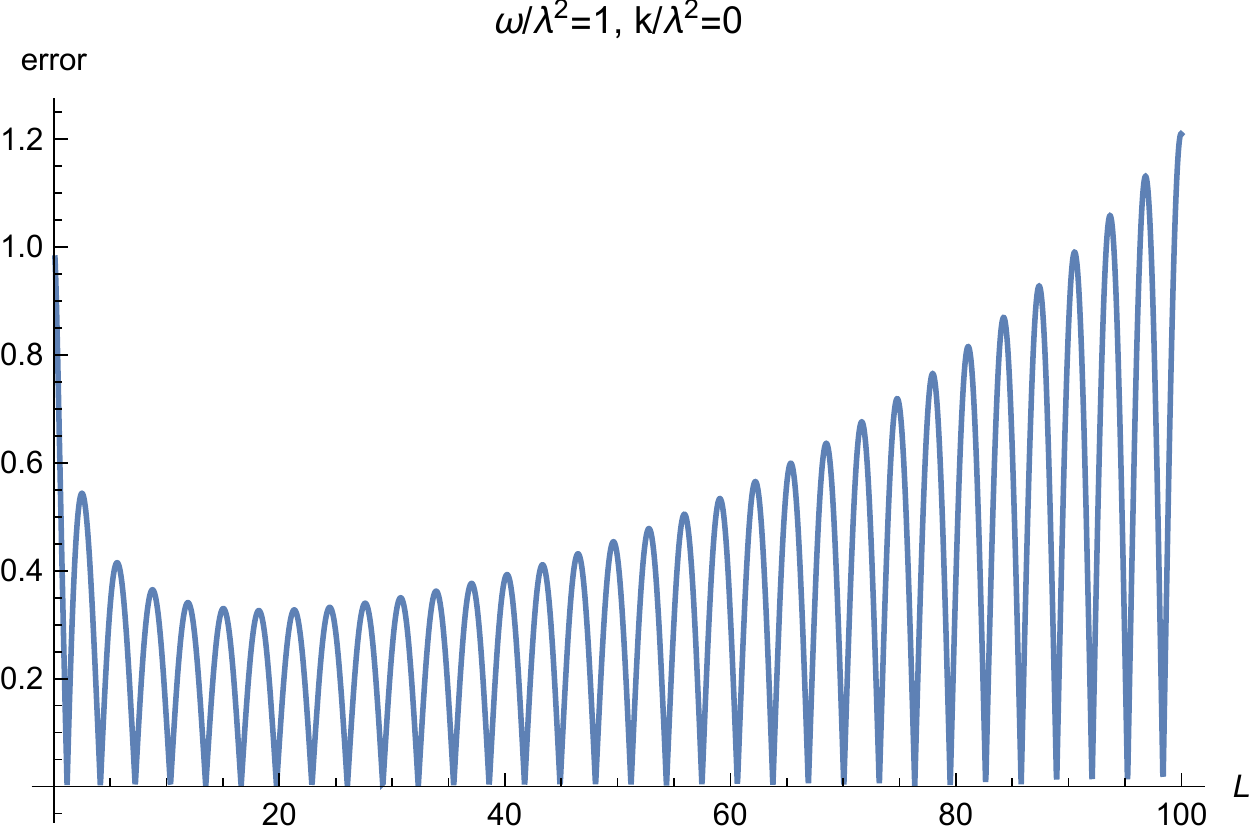}\includegraphics[width=0.33\textwidth]{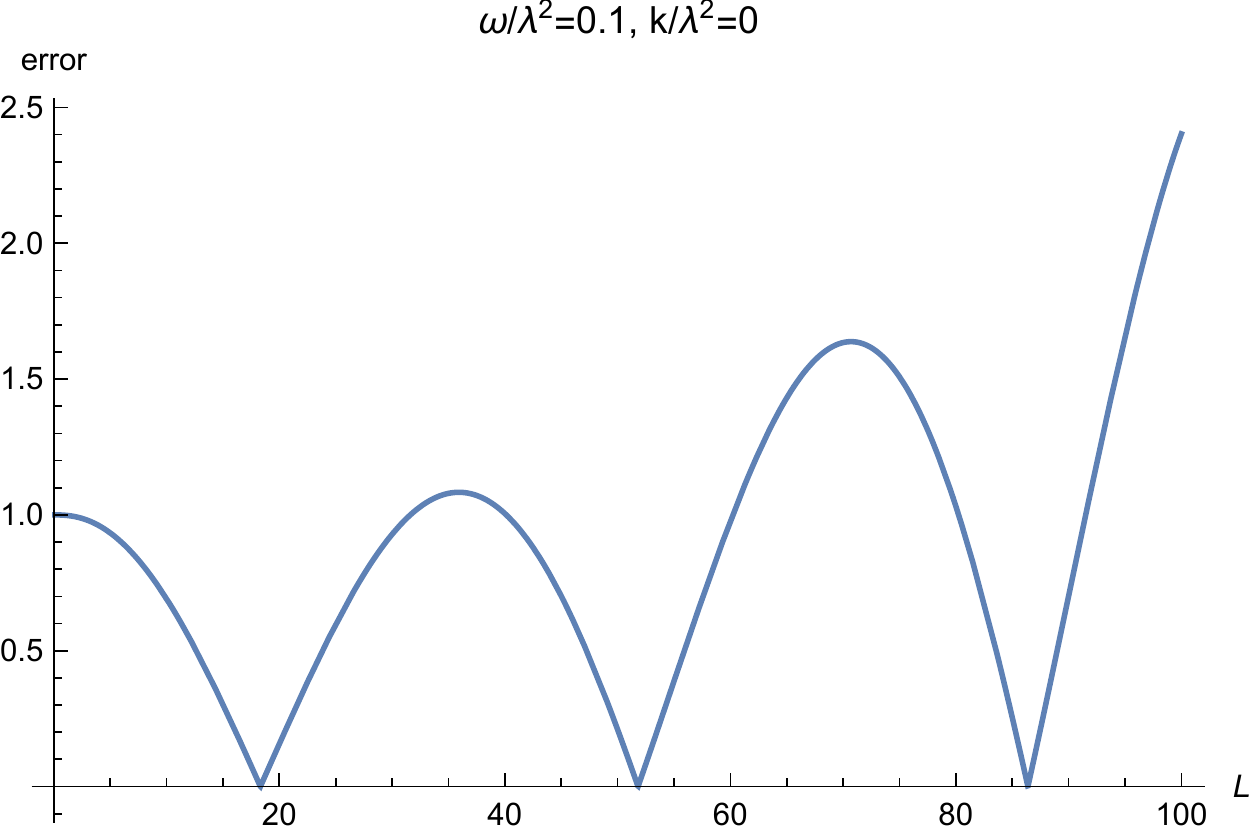}
\par\end{centering}

\caption{The relative difference between $G_{L}$ and the quenched result as
a function of $L$ for different frequencies and momenta. Left: for
a point in the $(\omega,k)$ plane which is outside of the shaded
region of Fig. \ref{fig:convergence} the ``error'' goes to zero
for large $L$. Middle: inside the shaded region the error is oscillating
with diverging amplitude as $L$ goes to larger values. However, for
larger values of $\omega$ there is an intermediate range of $L$,
where the relative error is smaller than 0.3. Therefore the quenched
approximation is qualitatively correct. Right: for small $\omega$ the
amplitude of the oscillation is large. \label{fig:error}}
\end{figure}

\end{document}